\documentclass[aps,showpacs,twocolumn,floatfix,superscriptaddress,nofootinbib,preprintnumbers,preprint]{ptptex}
\usepackage[dvips]{graphicx}
\usepackage{color}
\usepackage{amssymb,amsfonts,amsmath,longtable}

\preprintnumber{STUPP-07-194,\\ IFIC/08-15,\\ FTUV-08-0305}

\title{%
  Model building by coset space dimensional reduction scheme using
  ten-dimensional coset spaces
}

\author{Toshifumi~Jittoh,Masafumi~Koike,Takaaki~Nomura,Joe~Sato,Takashi~Shimomura*}
\inst{
  Saitama University,
  Shimo-Okubo,
  Sakura-ku,
  Saitama 355-8570, Japan;
  Departament de F\'{i}sica Te\`{o}rica and IFIC,
  Universitat de Val\`{e}ncia-CSIC,
  E-46100 Burjassot, Val\`{e}ncia, Spain*
}

\date{\today}

\abst{
\noindent
We investigate the gauge-Higgs unification models within the scheme of
the coset space dimensional reduction, beginning with a gauge theory
in a fourteen-dimensional spacetime where extra-dimensional space
has the structure of a ten-dimensional compact coset space.
We found seventeen phenomenologically acceptable models through an
exhaustive search for the candidates of the coset spaces, the gauge
group in fourteen dimension, and fermion representation.
Of the seventeen, ten models led to $\mathrm{SO}(10) (\times
\mathrm{U}(1))$ GUT-like models after dimensional reduction, three
models led to $\mathrm{SU}(5) \times \mathrm{U}(1)$ GUT-like
models, and four to $\mathrm{SU}(3) \times \mathrm{SU}(2)
\times \mathrm{U}(1) \times \mathrm{U}(1)$ Standard-Model-like
models.
The combinations of the coset space, the gauge group in the
fourteen-dimensional spacetime, and the representation of the
fermion contents of such models are listed.
}

\begin{document}

\maketitle

\section{Introduction}
\label{sec:Intro}

The Standard Model (SM) has described the
interactions of the elementary particles successfully.
In this model, the Higgs scalar plays an essential role in the
mechanism of spontaneous breaking of the gauge symmetry from
$\mathrm{SU}(3)_{\textrm{C}} \times \mathrm{SU}(2)_{\textrm{L}} \times
\mathrm{U}(1)_{Y}$ down to $\mathrm{SU}(3)_{\textrm{C}}
\times \mathrm{U}(1)_{\textrm{em}}$, giving masses to the elementary
particles. 
Nevertheless, the Higgs particle itself is still undiscovered.
Not only is it the last frontier of the SM, it will also provide the
key clue to the physics beyond the SM, since the SM does not address
even the most fundamental nature of the Higgs particle, such as its
mass and the self-coupling constants.

The gauge-Higgs unification is one of attractive approaches to the
physics beyond the SM in this regard~\cite{%
  Manton:1979kb,%
  Fairlie:1979at,%
  Fairlie:1979zy%
} (for recent approaches, see Refs.~\cite{%
  Hall:2001zb,%
  Burdman:2002se,%
  Gogoladze:2003ci,%
  Scrucca:2003ut,%
  Haba:2004qf,%
  Biggio:2004kr,%
  Maru:2004,%
  Haba:2005kc,%
  Hosotani:2006qp,%
  Sakamoto:2006wf,%
  Maru:2006,%
  Hosotani:2007qw,%
  Sakamura:2007qz,%
  Medina:2007hz,%
  Adachi:2007tj,%
  Gogoladze:2007ey%
}).
In this approach, the Higgs sector is embraced into the gauge
interactions in the spacetime with dimensions larger than four, where
the extra-dimensional space is compactified to a small scale to
reproduce the four-dimensional spacetime.
The scalar particles originate from the extra-dimensional components
of the gauge field and part of the fundamental properties of Higgs
scalar is determined from the gauge interactions.

We consider this approach in the framework of coset space dimensional
reduction (CSDR)~\cite{Kapetanakis:1992hf} (for recent approaces, see
Refs.~\cite{%
  Manousselis:2004xd,%
  Chatzistavrakidis:2007by,%
  Chatzistavrakidis:2007pp%
}).
This framework introduces a compact extra-dimensional space which has
the structure of a coset of Lie groups, and identifies the gauge
transformation as the translation within the extra-dimensional space.
This identification determines both the gauge symmetry and the
particle contents of the four-dimensional theory.

Gauge theories in six- and ten-dimensional spacetime are well
investigated.  No known model, however, reproduced the particle
content of the SM or Grand Unified Theory (GUT)~\cite{%
  Manton:1979kb,%
  Kapetanakis:1992hf,%
  10dim-Model:F2,%
  10dim-Model:D3,%
  10dim-Model:K4,%
  10dim-Model:N5,%
  10dim-Model:K6,%
  10dim-Model:K12,%
  10dim-Model:D14,%
  10dim-Model:B%
}.
The difficulty arises due to the insufficient dimensionality of the
vector and the spinor representations of the rotational group of the
extra-dimensional space, to one of which all the scalars and fermions
need to belong.
One of the promising ways to overcome this difficulty is to increase
the dimensionality of extra-dimensional space.
The higher-dimensional models enlarge these representations and thus
enriches the particle contents, which hopefully include particle
contents of the SM or GUT.
Another merit of such models is the increase of candidates of the
coset space and thus of the gauge group.

In determining the dimensionality of the extra-dimension, we pay
special attention to the chiral structure of the SM and GUT.
The representation of the gauge group in higher-dimension needs to be
complex in general to induce a chiral four-dimensional theory.
The exception is the case where the dimensionality is $4n + 2$.
This choice allows to start from vector-like representations, leaving
larger opportunity to search for an acceptable model.
Therefore we investigate fourteen-dimensional spacetime and search for
GUT, GUT-like model, the SM and the SM-like model.
A fourteen-dimensional theory is studied in terms of a bosonic string 
theory in a twenty-six dimensional spacetime \cite{Gavrilik}, giving another motivation 
of the models in the present work.

In this paper, we search for gauge theories in fourteen-dimensional
spacetime which leads to a phenomenologically acceptable model.
We exhaustively determined the coset spaces and the gauge groups.
The scalar contents are completely determined for each case and the
fermion contents are searched.
Here we consider the dimensions of fermion representations less than 1025 
since even larger representations yield numerous higher dimensional representations 
of fermion, under the gauge group of the SM or GUTs, in the four-dimensions.

This paper is organized as follows.
In Sec.~\ref{CSDR}, we give a brief review of the scheme of CSDR.
In Sec.~\ref{sec:candidates}, we consider the candidates of the
theories which lead to the phenomenologically acceptable models after
the dimensional reduction.
We summarize our results in Sec.~\ref{sec:summary}.

\section{The scheme of coset space dimensional reduction}
\label{CSDR}
In this section, we recapitulate the scheme of the coset space
dimensional reduction (CSDR) and the construction of the
four-dimensional theory by CSDR~\cite{Kapetanakis:1992hf}.

We begin with a gauge theory with a gauge group $G$ defined on a
$D$-dimensional spacetime $M^{D}$.
The spacetime $M^{D}$ is assumed to be a direct product of the
four-dimensional spacetime $M^{4}$ and a compact coset space $S/R$
such that $M^{D} = M^{4} \times S/R$, where $S$ is a compact Lie group
and $R$ is a Lie subgroup of $S$.
The dimension of the coset space $S/R$ is thus $d \equiv D - 4$, 
implying $\mathrm{dim} \, S - \dim \, R = d$.
This assumption on the structure of extra-dimensional space requires
the group $R$ to be embedded into the group $\mathrm{SO}(d)$, which is
a subgroup of the Lorentz group $\mathrm{SO}(1, D - 1)$.
Let us denote the coordinates of $M^{D}$ by $X^{M} = (x^{\mu},
y^{\alpha})$, where $x^{\mu}$ and $y^{\alpha}$ are coordinates of
$M^{4}$ and $S/R$, respectively.
The spacetime index $M$ runs over $\mu \in \{0, 1, 2, 3 \}$ and
$\alpha \in \{4, 5 , \cdots, D - 1 \}$.
We define the vielbein ${e_{M}}^{A}$ which relates the metric of the
manifold $M^{D}$ (the bulk spacetime), denoted by $g_{MN}(X)$, and
that of the tangent space $T_{X}M^{D}$ (the local Lorentz frame),
denoted by $h_{AB}(X)$, as $g_{MN} = {e_{M}}^{A} {e_{N}}^{B} h_{AB}$.
Here $A = (\mu, a)$, where $a \in \{ 4, \cdots, D \}$, is the index
for the coordinates of $T_{X}M^{D}$.
We conventionally use %
$\mu, \nu, \lambda, \cdots$ to denote the indices for $M^{4}$; %
$\alpha, \beta, \gamma, \cdots$ for the coset space $S/R$; %
$a, b, c, \cdots$ for the algebra of the group $S/R$; %
$M, N, \cdots$ for $(\mu, \alpha)$; %
and $A, B$ for $(\mu, a)$.
We introduce a gauge field $A_{M}(x, y) = (A_{\mu}(x,
y), A_{\alpha}(x, y))$, which belongs to the adjoint representation of
the gauge group $G$, and fermions $\psi(x, y)$, which lies 
in a representation $F$ of $G$.
The action $S$ of this theory is given by
%
\begin{equation}
\begin{split}
  S
  =
  &
  \int d^{D} X \sqrt{-g}
  \\
  \times
  \Bigl(
  &
    - \frac{1}{8} g^{MN} g^{KL} \,
      \mathrm{Tr} \, F_{MK}(X) F_{NL}(X)
  \\ &
    + \frac{1}{2} i \bar{\psi}(X) \Gamma^{A} {e_{A}}^{M} D_{M} \psi (X)
  \Bigr),
\end{split}
\label{eq:D-dim-action}
\end{equation} 
%
where
$g = \det g_{MN}$,
$F_{MN}(X) = \partial_{M} A_{N}(X) - \partial_{N}
A_{M}(X) - [A_{M}(X), A_{N}(X)]$ is the field strength,
$D_{M}$ is the covariant derivative on $M^{D}$,
and $\Gamma^{A}$ is the generators of the $D$-dimensional
Clifford algebra.

The extra-dimensional space $S/R$ admits $S$ as an isometric
transformation group, and we impose on $A_{M}(X)$ and $\psi(X)$ the
following symmetry under this transformation in order to carry out the
dimensional reduction \cite{Forgacs:1979zs,Symm-con:W13,Symm-con:J14,Symm-con:O15,Symm-con:Y16,Symm-con:Y16-2}.
Consider a coordinate transformation which acts trivially on $x$ and
gives rise to a $S$-transformation on $y$ as
\begin{equation}
  \label{cotra}
  (x, y) \rightarrow (x, sy), 
\end{equation}
where $s \in S$.
We require that this coordinate transformation Eq.~(\ref{cotra})
should be compensated by a gauge transformation.
This symmetry, connecting nontrivially the coordinate 
and gauge transformation, requires $R$ to be embedded into $G$. 
The symmetry further leads to the following set of
the symmetric condition on the fields:
\begin{subequations}
\begin{align}
  \label{g4cond}
  A_{\mu}(x, y) & = g(y;s) A_{\mu}(x, s^{-1}y) g^{-1}(y;s),
  \\
  \label{gexcond}
  A_{\alpha}(x, y)
  & = g(y; s) {J_{\alpha}}^{\beta} A_{\beta}(x, s^{-1}y) g^{-1}(y;s)
  \nonumber \\
  & \quad
      + g(y; s) \partial_{\alpha} g^{-1}(y; s),
  \\
  \label{psicond}
  \psi (x, y)
  & = f(y; s) \Omega \psi (x, s^{-1}y),
\end{align}
\label{eq:S-symmetry-cond}%
\end{subequations}
where $g(y;s)$ and $f(y;s)$ are gauge transformations in the adjoint
representation and in the representation $F$, respectively,
and ${J_{\alpha}}^{\beta}$ and $\Omega$ are the rotation in the
tangent space for the vectors and spinors, respectively.
These conditions of Eq.~(\ref{eq:S-symmetry-cond}) 
make the $D$-dimensional Lagrangian invariant under the $S$-transformation 
of Eq.~(\ref{cotra}) and therefore independent of the coordinate $y$ of $S/R$.
The dimensional reduction is then carried out by integrating over the
coordinate $y$ to obtain the four-dimensional Lagrangian.
The four-dimensional theory consists of the gauge fields $A_{\mu}$,
fermions $\psi$, and in addition the scalars $\phi_{a} \equiv
{e_{a}}^{\alpha} A_{\alpha}$.
The gauge group reduces to a subgroup $H$ of the original gauge
group $G$.
The dimensional reduction under the symmetric condition Eq. (\ref{eq:S-symmetry-cond}) 
and the assumption $h^{AB} = \mathrm{diag}
\, (\eta^{\mu \nu}, -g^{ab})$, where $\eta_{\mu \nu} = \mathrm{diag} \,
(1, -1, -1, -1)$ and $g^{ab} = \mathrm{diag} \, (a_{1}, a_{2}, \cdots,
a_{d})$ with $a_{i}$'s being positive, leads to the four-dimensional
effective Lagrangian $L^{\textrm{eff}}$ given by
\begin{equation}
\begin{split}
  L^{\textrm{eff}}
  = &
  - \frac{1}{4} F^{t}_{\mu \nu} F^{t \mu \nu}
  + \frac{1}{2} (D_{\mu} \phi_{a} )^{t} (D^{\mu} \phi^{a} )^{t} 
  + V(\phi)
  \\ &
  + \frac{1}{2} i \bar{\psi} \Gamma^{\mu} D_{\mu} \psi
  + \frac{1}{2} i \bar{\psi} \Gamma^{a} {e_{a}}^{\alpha} D_{\alpha} \psi,
\end{split}
\label{eq:4D-lagrangian}
\end{equation}
where $t$ is the index for the generators of the gauge group $G$.
It is notable that the Lagrangian Eq.~(\ref{eq:4D-lagrangian}) includes
the scalar potential $V(\phi)$, which is completely determined by the
group structure as
\begin{eqnarray}
  V(\phi)
  &=&
  - \frac{1}{4} g^{ac} g^{bd} \nonumber 
  \\ & & \;
  \times
  \mathrm{Tr} \,
  \Bigl[
    \bigl( {f_{ab}}^{C} \phi_{C} - [ \phi_{a}, \phi_{b} ] \bigr)
    \bigl( {f_{cd}}^{D} \phi_{D} - [ \phi_{c}, \phi_{d} ] \bigr)
  \Bigr],
\end{eqnarray}
where $C$ and $D$ runs over the indices of the algebra of $S$,
and
${f_{ab}}^{C}$ is the structure constants of the algebra of $S$.
This potential may cause the spontaneous symmetry breaking, rendering
the final gauge group $K$ a subgroup of the group $H$.

The scheme of CSDR substantially constrains the four-dimensional gauge
group $H$ and its representations for the particle contents as shown
below.
First, the gauge group of the four-dimensional theory $H$ is
easily identified as
\begin{equation}
  H = C_{G}(R),
\label{eq:H-is-centralizer}
\end{equation}
where $C_{G}(R)$ denotes the centralizer of $R$ in $G$ \cite{Forgacs:1979zs}.
Note that this implies $G \supset H \times R$ up to the U(1) factors.
Secondly, the representations of $H$ for the Higgs fields are
specified by the following prescription.
Suppose that the adjoint representations of $R$ and $G$ are decomposed
according to the embeddings $S \supset R$ and $G \supset H \times R$
as
\begin{align}
  \mathrm{adj} \, S & = \mathrm{adj} \, R + \sum_{s} r_{s},
  \label{dec}
  \\
  \mathrm{adj} \, G
  & = (\mathrm{adj} \, H, 1)
    + (1, \mathrm{adj} \, R)
    + \sum_{g} (h_{g}, r_{g}),
  \label{eq:G-rep-decomposition}
\end{align}
where $r_{s}$s and $r_{g}$s denote representations of $R$, and
$h_{g}$s denote representations of $H$.
The representation of the scalar fields are $h_{g}$s whose
corresponding $r_{g}$s in the decomposition
Eq.~(\ref{eq:G-rep-decomposition}) are contained also in the set $\{ r_{s}\}$.
Thirdly, the representation of $H$ for the fermion fields are
determined as follows \cite{Manton:1981es}.
Let the group $R$ be embedded into the Lorentz group $\mathrm{SO}(d)$ in
such a way that the vector representation $d$ of $\mathrm{SO}(d)$ is
decomposed as 
\begin{equation}
d = \sum_{s} r_{s},
\label{decS}
\end{equation} 
where $r_{s}$ are the representations
obtained in the decomposition Eq.~(\ref{decS}).
This embedding specifies a decomposition of the spinor representation
$\sigma_{d}$ of $\mathrm{SO}(d)$ into irreducible representations
$\sigma_{i}$s of $R$ as
\begin{equation}
  \sigma_{d} = \sum_{i} \sigma_{i}.
\label{eq:sigmad-decomposition}
\end{equation}
Now the representations of $H$ for the four-dimensional fermions are
found by decomposing $F$ according to $G \supset H \times R$ as
\begin{equation}
  F = \sum_{f} (h_{f}, r_{f}).
\label{eq:F-decomposition-to-HxR}
\end{equation}
The representations of our interest are $h_{f}$s whose corresponding
$r_{f}$s are found in $\{ \sigma_{i} \}$ obtained in
Eq.~(\ref{eq:sigmad-decomposition}).
Note that a phenomenologically acceptable model needs chiral
fermions in the four 
dimensions  
as the SM does.
This is possible only when the coset space $S/R$ satisfies $\mathrm{rank} \, S = \mathrm{rank} \, R$,
according to the non-trivial result due to Bott~\cite{Bott}.
The chiral fermions are then obtained most straightforwardly 
when we introduce a Weyl fermion in $D=2n$ ($n=1, 2, \cdots $) dimensions 
and $F$ is a complex representation \cite{Chapline:1982wy,Chiral-con:K12,Chiral-con:S13,Chiral-con:C14}.
Interestingly, they can be obtained even if $F$ is real or pseudoreal
representation, provided
$D = 4n + 2$ \cite{Chapline:1982wy,Chiral-con:C14}.
The four-dimensional fermions are doubled in these cases, and these
extra fermions are eliminated by imposing the Majorana condition on
the Weyl fermions in $D=4n+2$ dimensions\cite{Chapline:1982wy,Chiral-con:C14}.
From this condition we get chiral fermions for
$D = 8n + 2$ ($8n + 6$) when $F$ is real (pseudoreal).
It is therefore interesting to consider $D =6, 10, 14, 18, \cdots$.

\section{The Search for acceptable candidates}
\label{sec:candidates}
In this section, we search for candidates of the coset space $S/R$,
the gauge group $G$, and its representation $F$ for fermions in the
spacetime of the dimensionality $D=14$ for phenomenologically
acceptable models based on CSDR scheme.
Such models should induce a four-dimensional theory that has a gauge
group $H \supset \mathrm{SU}(3) \times \mathrm{SU}(2) \times
\mathrm{U}(1)$, and accomodates chiral fermions contained in the
SM.
This requirement constrains the $D$, $S/R$, $G$, $F$, 
and the embedding of $R$ in $G$.

Number of dimensions $D$ should be $2n$ in order to give chiral
fermions in four 
dimensions.
We are particularly interested in the case of $D = 4n + 2$, where
chiral fermions can be obtained in four dimensions even if $F$ is real
or pseudoreal. 
The simplest cases of $D = 6$ and $10$ are well investigated. 
No known model, however, reproduced the particle contents of the SM or GUT. 
\cite{Manton:1979kb,Kapetanakis:1992hf,10dim-Model:F2,10dim-Model:D3,10dim-Model:K4,10dim-Model:N5,10dim-Model:K6,10dim-Model:K12,10dim-Model:D14,10dim-Model:B}.
This is due to the small dimensionality of the vector and spinor representations of $\textrm{SO}(d)$.
It is difficult when $d=2$ and $6$ to match 
$r_s$s from $\textrm{SO}(d)$ vector and $\sigma_i$s from $\textrm{SO}(d)$ spinor 
with $r_g$s from $\mathrm{adj} \, G$ and $r_f$s from $F$, respectively (see Eqs.~(\ref{decS})-(\ref{eq:F-decomposition-to-HxR})).
We consider a higher-dimensional spacetime to enlarge the dimensionality of 
$\textrm{SO}(d)$ vector and spinor representations.  
More $r_g$s and $r_f$s will satisfy the matching prescription, and hence richer particle contents are obtained.
Another merit of higher-dimensional spacetime is the increase of candidates of the coset space and thus of the gauge group.
We thus investigate next smallest dimensionality of $D=4n+2$, which is $D = 14$.

Coset space $S/R$ of our interest should have dimension $d = D - 4 =
10$, implying $\mathrm{dim} \, S - \dim \, R = 10$, and should satisfy
$\mathrm{rank} \, S = \mathrm{rank} \, R$ to generate chiral fermions
in four dimensions \cite{Bott}.
These conditions limit the possible $S/R$ to the coset spaces collected in
Table~\ref{table_10dim_coset_spaces}.
There the correspondence between the subgroup of $R$ and the
subgroup of $S$ is clarified by the brackets in $R$.
For example, the coset space (2) suggests direct sum of
$\mathrm{SO}(7)/\mathrm{SO}(6)$ and $\mathrm{Sp}(4)/[\mathrm{SU}(2)
\times \mathrm{SU}(2)]$.
The factor of $R$ with subscript ``max'' indicates that this factor is
a maximal regular subalgebra of $S$.
For example, the coset (20) in Table~\ref{table_10dim_coset_spaces}
indicates that $[\mathrm{SU}(2) \times \mathrm{U}(1) ]_{\textrm{max}}$
is the maximal regular subgroup of $\mathrm{Sp}(4)$. 
We show the embedding of $R$ in SO(10) in
Table~\ref{so10_vector_and_spinor_with_u1u1}.
The representations of $r_{s}$ in Eq.~(\ref{decS}) and $\sigma_{i}$ in
Eq.~(\ref{eq:sigmad-decomposition}) are listed in the columns of
``Branches of $\mathbf{10}$'' and ``Branches of $\mathbf{16}$'',
respectively.
\begingroup


\begin{table}
  \caption{%
    A complete list of ten-dimensional coset spaces $S/R$ with
    $\mathrm{rank} \, S = \mathrm{rank} \, R$.
    The brackets in $R$ clarifies the
    correspondence between the subgroup of $R$ and the subgroup of $S$.
    The factor of $R$ with subscript ``max'' indicates that this factor is
    a maximal regular subalgebra of $S$.
  }
    \label{table_10dim_coset_spaces}
\begin{center}
\let\tabularsize\small
\begin{tabular}{ll}
  \hline
  No. & $S/R$
  \\ \hline \hline
  (1)
  & $\mathrm{SO}(11) / \mathrm{SO}(10)$
  \\ \hline
  (2)
  & $\mathrm{SO}(7) \times \mathrm{Sp}(4)
     / \mathrm{SO}(6) \times [\mathrm{SU}(2) \times \mathrm{SU}(2)]$
  \\ \hline
  (3)
  & $\mathrm{G}_{2} \times \mathrm{Sp}(4)
     / \mathrm{SU}(3) \times [\mathrm{SU}(2) \times \mathrm{SU}(2)]$
  \\ \hline
  (4)
  & $\mathrm{SU}(6)
     / \mathrm{SU}(5) \times \mathrm{U}(1)$
  \\ \hline
  (5)
  & $\mathrm{SO}(9) \times \mathrm{SU}(2)
     / \mathrm{SO}(8) \times \mathrm{U}(1)$
  \\ \hline
  (6)
  & $\mathrm{SO}(7) \times \mathrm{SU}(3)
     / \mathrm{SO}(6) \times [\mathrm{SU}(2) \times \mathrm{U}(1)]$
  \\ \hline
  (7)
  & $\mathrm{SU}(4) \times \mathrm{Sp}(4)
     / [\mathrm{SU}(3) \times \mathrm{U}(1)]
       \times [\mathrm{SU}(2) \times \mathrm{SU}(2)]$
  \\ \hline
  (8)
  & $(\mathrm{Sp}(4))^{2} \times \mathrm{SU}(2)
     / [\mathrm{SU}(2) \times \mathrm{SU}(2)]^{2} \times \mathrm{U}(1)$
  \\ \hline
  (9)
  & $\mathrm{G}_{2} \times \mathrm{SU}(3)
     / \mathrm{SU}(3) \times [\mathrm{SU}(2) \times \mathrm{U}(1)]$
  \\ \hline
  (10)
  & $\mathrm{Sp}(4) \times \mathrm{Sp}(4)
     / [\mathrm{SU}(2) \times \mathrm{U}(1)]_{\textrm{max}}
       \times [\mathrm{SU}(2) \times \mathrm{SU}(2)]$
  \\ \hline
  (11)
  & $\mathrm{Sp}(4) \times \mathrm{Sp}(4)
     / [\mathrm{SU}(2) \times \mathrm{U}(1)]_{\textrm{non-max}}
       \times [\mathrm{SU}(2) \times \mathrm{SU}(2)]$
  \\ \hline
  (12)
  & $\mathrm{Sp}(6) \times \mathrm{SU}(2)
     / [\mathrm{Sp}(4) \times \mathrm{SU}(2)] \times \mathrm{U}(1)$
  \\ \hline
  (13)
  & $\mathrm{G}_{2} \times \mathrm{SU}(2)
     / \mathrm{SU}(2) \times \mathrm{SU}(2) \times \mathrm{U}(1)$
  \\ \hline
  (14)
  & $\mathrm{Sp}(6)
     / \mathrm{Sp}(4) \times \mathrm{U}(1)$
  \\ \hline
  (15)
  & $\mathrm{G}_{2}
     / \mathrm{SU}(2) \times \mathrm{U}(1)$
  \\ \hline
  (16)
  & $\mathrm{Sp}(4) \times \mathrm{SU}(3) \times \mathrm{SU}(2)
     / [\mathrm{SU}(2) \times \mathrm{SU}(2)]
       \times [\mathrm{SU}(2) \times \mathrm{U}(1)]
       \times \mathrm{U}(1)$
  \\ \hline
  (17)
  & $\mathrm{SU}(4) \times \mathrm{SU}(3)
     / [\mathrm{SU}(3) \times \mathrm{U}(1)]
       \times [\mathrm{SU}(2) \times \mathrm{U}(1)]$
  \\ \hline
  (18)
  & $\mathrm{SO}(7) \times (\mathrm{SU}(2))^{2}
     / \mathrm{SO}(6) \times (\mathrm{U}(1))^{2}$
  \\ \hline
  (19)
  & $\mathrm{SU}(5) \times \mathrm{SU}(2)
     / [\mathrm{SU}(4) \times \mathrm{U}(1)] \times \mathrm{U}(1)$
  \\ \hline
  (20)
  & $\mathrm{Sp}(4) \times \mathrm{SU}(3)
     / [\mathrm{SU}(2) \times \mathrm{U}(1)]_{\textrm{max}}
       \times [\mathrm{SU}(2) \times \mathrm{U}(1)]$
  \\ \hline
  (21)
  & $\mathrm{Sp}(4) \times \mathrm{SU}(3)
     / [\mathrm{SU}(2) \times \mathrm{U}(1)]_{\textrm{non-max}}
       \times [\mathrm{SU}(2) \times \mathrm{U}(1)]$
  \\ \hline
  (22)
  & $\mathrm{SU}(3) \times \mathrm{Sp}(4)
     / [\mathrm{U}(1) \times \mathrm{U}(1)]
       \times [\mathrm{SU}(2) \times \mathrm{SU}(2)]$
  \\ \hline
  (23)
  & $\mathrm{SU}(4) \times \mathrm{SU}(2)
    / \mathrm{SU}(2) \times \mathrm{SU}(2) \times \mathrm{U}(1)
      \times \mathrm{U}(1)$
  \\ \hline
  (24)
  & $\mathrm{G}_{2} \times (\mathrm{SU}(2))^{2}
     / \mathrm{SU}(3) \times (\mathrm{U}(1))^{2}$ 
  \\ \hline
  (25)
  & $\mathrm{SU}(4)
     / \mathrm{SU}(2) \times \mathrm{U}(1) \times \mathrm{U}(1)$ 
  \\ \hline
  (26)
  & $\mathrm{Sp}(4) \times (\mathrm{SU}(2))^{3}
     / [\mathrm{SU}(2) \times \mathrm{SU}(2)] \times (\mathrm{U}(1))^{3} $ 
  \\ \hline
  (27)
  & $(\mathrm{SU}(3))^{2} \times \mathrm{SU}(2)
     / [\mathrm{SU}(2) \times \mathrm{U}(1)]^{2} \times \mathrm{U}(1)$
  \\ \hline
  (28)
  & $\mathrm{SU}(4) \times (\mathrm{SU}(2))^{2}
     / [\mathrm{SU}(3) \times \mathrm{U}(1)] \times (\mathrm{U}(1))^{2}$
  \\ \hline
  (29)
  & $\mathrm{Sp}(4) \times (\mathrm{SU}(2))^{2}
     / [\mathrm{SU}(2) \times \mathrm{U}(1)]_{\textrm{max}}
       \times (\mathrm{U}(1))^{2}$
  \\ \hline
  (30)
  & $\mathrm{Sp}(4) \times (\mathrm{SU}(2))^{2}
     / [\mathrm{SU}(2) \times \mathrm{U}(1)]_{\textrm{non-max}}
       \times (\mathrm{U}(1))^{2}$
  \\ \hline
  (31)
  & $\mathrm{SU}(3) \times \mathrm{SU}(3)
     / [\mathrm{U}(1) \times \mathrm{U}(1)]
       \times [\mathrm{SU}(2) \times \mathrm{U}(1)]$
  \\ \hline
  (32)
  & $\mathrm{Sp}(4) \times \mathrm{SU}(2)
     / [\mathrm{U}(1) \times \mathrm{U}(1)] \times \mathrm{U}(1)$
  \\ \hline
  (33)
  & $\mathrm{SU}(3) \times (\mathrm{SU}(2))^{3}
     / [\mathrm{SU}(2) \times \mathrm{U}(1)] \times (\mathrm{U}(1))^{3}$
  \\ \hline
  (34)
  & $(\mathrm{SU}(2) / \mathrm{U}(1))^{5}$
  \\ \hline
  (35)
  & $\mathrm{SU}(3) \times (\mathrm{SU}(2))^{2}
    / [\mathrm{U}(1) \times \mathrm{U}(1)] \times (\mathrm{U}(1))^{2}$
  \\ \hline
\end{tabular}
\end{center}
\end{table}
\endgroup
\begingroup
\begin{table}
  \caption{%
  {\footnotesize
    The decompositions of the vector representation $\mathbf{10}$
    and the spinor representation $\mathbf{16}$ of $\mathrm{SO}(10)$
    under $R$'s which are listed in Table~\ref{table_10dim_coset_spaces}
    and have two or less $\mathrm{U}(1)$ factors.
    The representations of $r_{s}$ in Eq.~(\ref{decS}) and $\sigma_{i}$ 
    in Eq.~(\ref{eq:sigmad-decomposition}) 
    are listed in the columns of ``Branches of $\mathbf{10}$'' and 
    ``Branches of $\mathbf{16}$'', respectively. 
    The $\mathrm{U}(1)$ charges for the cosets (16) -- (35) have a
    freedom of retaking the linear combination.}
  }
  \label{so10_vector_and_spinor_with_u1u1}
\begin{center}
\let\tabularsize\scriptsize

  \renewcommand{\arraystretch}{1.2}
\begin{tabular}{llll}
  \hline
  $S/R$
  &
  & Branches of $\mathbf{10}$
  & Branches of $\mathbf{16}$
  \\ \hline \hline
  (1) 
  & $\mathrm{SO}(10)$
  & $\mathbf{10}$ 
  & $\mathbf{16}$ 
  \\
  \hline
   (2) 
  & $(\mathrm{SO}(6), \mathrm{SU}(2), \mathrm{SU}(2))$
  & $(\mathbf{6},\mathbf{1},\mathbf{1}),(\mathbf{1},\mathbf{2},\mathbf{2})$ 
  & $(\mathbf{4},\mathbf{2},\mathbf{1}),(\bar{\mathbf{4}},\mathbf{1},\mathbf{2})$ 
  \\
  \hline
  (3)
  & $ (\mathrm{SU}(3), \mathrm{SU}(2), \mathrm{SU}(2))$
  & $ (\mathbf{3},\mathbf{1},\mathbf{1}),
	  (\bar{\mathbf{3}},\mathbf{1},\mathbf{1}), (\mathbf{1},\mathbf{2},\mathbf{2})$ 
  & $ (\mathbf{3},\mathbf{2},\mathbf{1}), (\bar{\mathbf{3}},\mathbf{1},\mathbf{2}),
      (\mathbf{1},\mathbf{2},\mathbf{1}), (\mathbf{1},\mathbf{1},\mathbf{2})$ 
  \\
  \hline
  (4)
  & $\mathrm{SU}(5)(\mathrm{U}(1))$
  & $\mathbf{5} (6)$,   
    $\bar{\mathbf{5}} (-6)$  
  & $\mathbf{1} (-15)$,   
    $\bar{\mathbf{5}} (9)$,   
    $\mathbf{10} (-3)$  
  \\ 
  \hline
  (5)
  & $\mathrm{SO}(8)(\mathrm{U}(1))$
  & $\mathbf{8}_{\textrm{v}} (0)$,   
    $\mathbf{1} (2)$,   
    $\mathbf{1} (-2)$  
  & $\mathbf{8}_{\textrm{s}} (-1)$,   
    $\mathbf{8}_{\textrm{c}} (1)$,   
  \\ 
  \hline
  (6)
  & $(\mathrm{SO}(6), \mathrm{SU}(2))(\mathrm{U}(1))$
  & $(\mathbf{6}, \mathbf{1}) (0)$,   
    $(\mathbf{1}, \mathbf{2}) (3)$,   
    $(\mathbf{1}, \mathbf{2}) (-3)$  
  & $(\mathbf{4}, \mathbf{2}) (0)$,   
    $(\bar{\mathbf{4}}, \mathbf{1}) (3)$,   
    $(\bar{\mathbf{4}}, \mathbf{1}) (-3)$  
  \\ 
  \hline
  (7)
  & $(\mathrm{SU}(3), \mathrm{SU}(2), \mathrm{SU}(2))(\mathrm{U}(1))$
  & $(\mathbf{3}, \mathbf{1}, \mathbf{1}) (-4)$,   
    $(\bar{\mathbf{3}}, \mathbf{1}, \mathbf{1}) (4)$,   
  & $(\mathbf{3}, \mathbf{1}, \mathbf{2}) (2)$,   
    $(\bar{\mathbf{3}}, \mathbf{2}, \mathbf{1}) (-2)$,   
  \\ 
  & 
  & $(\mathbf{1}, \mathbf{2}, \mathbf{2}) (0)$  
  & $(\mathbf{1}, \mathbf{1}, \mathbf{2}) (-6)$,   
    $(\mathbf{1}, \mathbf{2}, \mathbf{1}) (6)$  
  \\ 
  \hline
  (8)
  & $(\mathrm{SU}(2), \mathrm{SU}(2), \mathrm{SU}(2), \mathrm{SU}(2))(\mathrm{U}(1))$
  & $(\mathbf{2}, \mathbf{2}, \mathbf{1}, \mathbf{1}) (0)$,   
    $(\mathbf{1}, \mathbf{1}, \mathbf{2}, \mathbf{2}) (0)$,   
  & $(\mathbf{2}, \mathbf{1}, \mathbf{1}, \mathbf{2}) (1)$,   
    $(\mathbf{1}, \mathbf{2}, \mathbf{1}, \mathbf{2}) (-1)$,   
  \\ 
  & 
  & $(\mathbf{1}, \mathbf{1}, \mathbf{1}, \mathbf{1}) (2)$,   
    $(\mathbf{1}, \mathbf{1}, \mathbf{1}, \mathbf{1}) (-2)$  
  & $(\mathbf{2}, \mathbf{1}, \mathbf{2}, \mathbf{1}) (-1)$,   
    $(\mathbf{1}, \mathbf{2}, \mathbf{2}, \mathbf{1}) (1)$  
  \\ 
  \hline
  (9)
  & $(\mathrm{SU}(3), \mathrm{SU}(2))(\mathrm{U}(1))$
  & $(\mathbf{3}, \mathbf{1}) (0)$,   
    $(\bar{\mathbf{3}}, \mathbf{1}) (0)$,   
    $(\mathbf{1}, \mathbf{2}) (3)$,   
  & $(\mathbf{3}, \mathbf{2}) (0)$,   
    $(\bar{\mathbf{3}}, \mathbf{1}) (3)$,   
    $(\bar{\mathbf{3}}, \mathbf{1}) (-3)$,   
  \\ 
  & 
  & $(\mathbf{1}, \mathbf{2}) (-3)$  
  & $(\mathbf{1}, \mathbf{2}) (0)$,   
    $(\mathbf{1}, \mathbf{1}) (3)$,   
    $(\mathbf{1}, \mathbf{1}) (-3)$  
  \\ 
  \hline
  (10)
  & $(\mathrm{SU}(2), \mathrm{SU}(2), \mathrm{SU}(2))(\mathrm{U}(1))$
  & $(\mathbf{2}, \mathbf{2}, \mathbf{1}) (0)$,   
    $(\mathbf{1}, \mathbf{1}, \mathbf{3}) (2)$,   
  & $(\mathbf{2}, \mathbf{1}, \mathbf{3}) (-1)$,   
    $(\mathbf{1}, \mathbf{2}, \mathbf{3}) (1)$,   
  \\ 
  & 
  & $(\mathbf{1}, \mathbf{1}, \mathbf{3}) (-2)$  
  & $(\mathbf{1}, \mathbf{2}, \mathbf{1}) (3)$,   
    $(\mathbf{2}, \mathbf{1}, \mathbf{1}) (-3)$  
  \\ 
  \hline
  (11)
  & $(\mathrm{SU}(2), \mathrm{SU}(2), \mathrm{SU}(2))(\mathrm{U}(1))$
  & $(\mathbf{2}, \mathbf{2}, \mathbf{1}) (0)$,   
    $(\mathbf{1}, \mathbf{1}, \mathbf{2}) (1)$,   
  & $(\mathbf{1}, \mathbf{2}, \mathbf{2}) (-1)$,   
    $(\mathbf{1}, \mathbf{2}, \mathbf{1}) (0)$,   
  \\ 
  & 
  & $(\mathbf{1}, \mathbf{1}, \mathbf{2}) (-1)$,   
    $(\mathbf{1}, \mathbf{1}, \mathbf{1}) (2)$  
  & $(\mathbf{1}, \mathbf{2}, \mathbf{1}) (2)$,   
    $(\mathbf{2}, \mathbf{1}, \mathbf{2}) (1)$  
  \\ 
  & 
  & $(\mathbf{1}, \mathbf{1}, \mathbf{1}) (-2)$  
  & $(\mathbf{2}, \mathbf{1}, \mathbf{1}) (0)$,   
    $(\mathbf{2}, \mathbf{1}, \mathbf{1}) (-2)$  
  \\ 
  \hline
  (12)
  & $(\mathrm{Sp}(4), \mathrm{SU}(2))(\mathrm{U}(1))$
  & $(\mathbf{4}, \mathbf{2}) (0)$,   
    $(\mathbf{1}, \mathbf{1}) (2)$,   
    $(\mathbf{1}, \mathbf{1}) (-2)$  
  & $(\mathbf{5}, \mathbf{1}) (-1)$,   
    $(\mathbf{1}, \mathbf{3}) (-1)$,   
    $(\mathbf{4}, \mathbf{2}) (1)$  
  \\ 
  \hline
  (13)
  & $(\mathrm{SU}(2), \mathrm{SU}(2))(\mathrm{U}(1))$
  & $(\mathbf{4}, \mathbf{2}) (0)$,   
    $(\mathbf{1}, \mathbf{1}) (2)$,   
    $(\mathbf{1}, \mathbf{1}) (-2)$  
  & $(\mathbf{4}, \mathbf{2}) (1)$,   
    $(\mathbf{5}, \mathbf{1}) (-1)$,   
    $(\mathbf{1}, \mathbf{3}) (-1)$  
  \\ 
  \hline
  (14)
  & $\mathrm{Sp}(4)(\mathrm{U}(1))$
  & $\mathbf{4} (1)$,   
    $\mathbf{4} (-1)$,   
    $\mathbf{1} (2)$,   
    $\mathbf{1} (-2)$  
  & $\mathbf{5} (1)$,   
    $\mathbf{4} (-2)$,   
    $\mathbf{4} (0)$,   
    $\mathbf{1} (3)$,   
    $\mathbf{1} (1)$,   
    $\mathbf{1} (-1)$,   
  \\ 
  \hline
  (15a)
  & $\mathrm{SU}(2)(\mathrm{U}(1))$
  & $\mathbf{2} (3)$,   
    $\mathbf{2} (-3)$,   
    $\mathbf{2} (1)$,   
    $\mathbf{2} (-1)$,   
  & $\mathbf{3} (1)$,   
    $\mathbf{2} (-4)$,   
    $\mathbf{2} (2)$,   
    $\mathbf{2} (-2)$,   
  \\ 
  & 
  & $\mathbf{1} (-2)$,   
    $\mathbf{1} (2)$  
  & $\mathbf{2} (0)$,   
    $\mathbf{1} (5)$,   
    $\mathbf{1} (3)$,   
    $\mathbf{1} (-3)$,  
    $\mathbf{1} (1)$,   
    $\mathbf{1} (-1)$  
  \\ 
  \hline
  (15b)
  & $\mathrm{SU}(2)(\mathrm{U}(1))$
  & $\mathbf{4} (1)$,   
    $\mathbf{4} (-1)$,   
    $\mathbf{1} (2)$,   
    $\mathbf{1} (-2)$  
  & $\mathbf{5} (1)$,   
    $\mathbf{4} (-2)$,   
    $\mathbf{4} (0)$,   
    $\mathbf{1} (3)$,   
    $\mathbf{1} (1)$,   
    $\mathbf{1} (-1)$,   
  \\ 
  \hline
  (16)
  & $(\mathrm{SU}(2), \mathrm{SU}(2), \mathrm{SU}(2))(\mathrm{U}(1), \mathrm{U}(1))$
  & $(\mathbf{2}, \mathbf{2}, \mathbf{1}) (0, 0)$,   
    $(\mathbf{1}, \mathbf{1}, \mathbf{2}) (3, 0)$,   
  & $(\mathbf{2}, \mathbf{1}, \mathbf{2}) (0, 1)$,   
    $(\mathbf{1}, \mathbf{2}, \mathbf{2}) (0, -1)$,   
  \\ 
  & 
  & $(\mathbf{1}, \mathbf{1}, \mathbf{2}) (-3, 0)$,   
    $(\mathbf{1}, \mathbf{1}, \mathbf{1}) (0, 2)$  
  & $(\mathbf{2}, \mathbf{1}, \mathbf{1}) (3, -1)$,   
    $(\mathbf{2}, \mathbf{1}, \mathbf{1}) (-3, -1)$  
  \\ 
  & 
  & $(\mathbf{1}, \mathbf{1}, \mathbf{1}) (0, -2)$  
  & $(\mathbf{1}, \mathbf{2}, \mathbf{1}) (3, 1)$,   
    $(\mathbf{1}, \mathbf{2}, \mathbf{1}) (-3, 1)$  
  \\ 
  \hline
  (17)
  & $(\mathrm{SU}(3), \mathrm{SU}(2))(\mathrm{U}(1), \mathrm{U}(1))$
  & $(\mathbf{3}, \mathbf{1}) (0, -4)$,   
    $(\bar{\mathbf{3}}, \mathbf{1}) (0, 4)$,   
  & $(\mathbf{3}, \mathbf{2}) (0, 2)$,   
    $(\bar{\mathbf{3}}, \mathbf{1}) (3, -2)$,   
    $(\bar{\mathbf{3}}, \mathbf{1}) (-3, -2)$,   
  \\ 
  & 
  & $(\mathbf{1}, \mathbf{2}) (3, 0)$,   
    $(\mathbf{1}, \mathbf{2}) (-3, 0)$  
  & $(\mathbf{1}, \mathbf{2}) (0, -6)$,  
    $(\mathbf{1}, \mathbf{1}) (3, 6)$,  
    $(\mathbf{1}, \mathbf{1}) (-3, 6)$  
  \\ 
  \hline
  (18)
  & $\mathrm{SO}(6)(\mathrm{U}(1), \mathrm{U}(1))$
  & $\mathbf{6} (0, 0)$,   
    $\mathbf{1} (2, 0)$,   
    $\mathbf{1} (-2, 0)$,   
  & $\mathbf{4} (1, -1)$,   
    $\mathbf{4} (-1, 1)$,   
    $\bar{\mathbf{4}} (1, 1)$,   
  \\ 
  & 
  & $\mathbf{1} (0, 2)$,   
    $\mathbf{1} (0, -2)$  
  & $\bar{\mathbf{4}} (-1, -1)$,   
  \\ 
  \hline
  (19)
  & $\mathrm{SU}(4)(\mathrm{U}(1), \mathrm{U}(1))$
  & $\mathbf{4} (0, -5)$,   
    $\bar{\mathbf{4}} (0, 5)$,   
    $\mathbf{1} (2, 0)$,   
  & $\mathbf{6} (-1, 0)$,   
    $\mathbf{4} (1, 5)$,   
    $\bar{\mathbf{4}} (1, -5)$,   
  \\ 
  & 
  & $\mathbf{1} (-2, 0)$  
  & $\mathbf{1} (-1, 10)$,   
    $\mathbf{1} (-1, -10)$  
  \\ 
  \hline
  (20)
  & $(\mathrm{SU}(2), \mathrm{SU}(2))(\mathrm{U}(1), \mathrm{U}(1))$
  & $(\mathbf{3}, \mathbf{1}) (0, 2)$,   
    $(\mathbf{3}, \mathbf{1}) (0, -2)$,   
  & $(\mathbf{3}, \mathbf{2}) (0, -1)$,   
    $(\mathbf{3}, \mathbf{1}) (3, 1)$,   
    $(\mathbf{3}, \mathbf{1}) (-3, 1)$,   
  \\ 
  & 
  & $(\mathbf{1}, \mathbf{2}) (3, 0)$,   
    $(\mathbf{1}, \mathbf{2}) (-3, 0)$  
  & 
    $(\mathbf{1}, \mathbf{2}) (0, 3)$,  
    $(\mathbf{1}, \mathbf{1}) (3, -3)$,   
    $(\mathbf{1}, \mathbf{1}) (-3, -3)$   
  \\ 
  \hline
  (21)
  & $(\mathrm{SU}(2), \mathrm{SU}(2))(\mathrm{U}(1), \mathrm{U}(1))$
  & $(\mathbf{2}, \mathbf{1}) (1, 0)$,   
    $(\mathbf{2}, \mathbf{1}) (-1, 0)$,   
  & $(\mathbf{2}, \mathbf{2}) (-1, 0)$,   
    $(\mathbf{1}, \mathbf{2}) (2, 0)$,   
    $(\mathbf{1}, \mathbf{2}) (0, 0)$,   
  \\ 
  & 
  & $(\mathbf{1}, \mathbf{2}) (0, 3)$,   
    $(\mathbf{1}, \mathbf{2}) (0, -3)$  
  & 
    $(\mathbf{2}, \mathbf{1}) (1, 3)$,  
    $(\mathbf{2}, \mathbf{1}) (1, -3)$,   
    $(\mathbf{1}, \mathbf{1}) (0, 3)$,  
  \\ 
  & 
  & $(\mathbf{1}, \mathbf{1}) (2, 0)$,   
    $(\mathbf{1}, \mathbf{1}) (-2, 0)$  
  & $(\mathbf{1}, \mathbf{1}) (0, -3)$,   
    $(\mathbf{1}, \mathbf{1}) (-2, 3)$,  
    $(\mathbf{1}, \mathbf{1}) (-2, -3)$,  
  \\ 
  \hline
  (22)
  & $(\mathrm{SU}(2), \mathrm{SU}(2))(\mathrm{U}(1), \mathrm{U}(1))$
  & $(\mathbf{2}, \mathbf{2}) (0, 0)$,   
    $(\mathbf{1}, \mathbf{1}) (a, c)$,   
  & $(\mathbf{2}, \mathbf{1}) (0, 0)$,   
    $(\mathbf{1}, \mathbf{2}) (0, 0)$,   
  \\ 
  & 
  & $(\mathbf{1}, \mathbf{1}) (b, d)$,   
    $(\mathbf{1}, \mathbf{1}) (-a, -c)$  
  & $(\mathbf{2}, \mathbf{1}) (b, d)$,   
    $(\mathbf{2}, \mathbf{1}) (a, c)$  
  \\ 
  & 
  & $(\mathbf{1}, \mathbf{1}) (-b, -d)$,   
  & $(\mathbf{2}, \mathbf{1}) (- a - b, - c - d)$,   
  \\ 
  & 
  & $(\mathbf{1}, \mathbf{1}) (a + b, c + d)$,   
  & $(\mathbf{1}, \mathbf{2}) (a + b, c + d)$,   
  \\ 
  & 
  & $(\mathbf{1}, \mathbf{1}) (- a - b, - c - d)$  
  & $(\mathbf{1}, \mathbf{2}) (-a, -c)$,   
    $(\mathbf{1}, \mathbf{2}) (-b, -d)$  
  \\ 
  \hline
\end{tabular}
\end{center}
\end{table}
\begin{table}
  \addtocounter{table}{-1}
  \caption{(Continued.)}
\begin{center}
 \let\tabularsize\scriptsize
  \renewcommand{\arraystretch}{1.2}
\begin{tabular}{llll}
  \hline
  $S/R$
  &
  & Branches of $\mathbf{10}$
  & Branches of $\mathbf{16}$
  \\ \hline \hline
  (23)
  & $(\mathrm{SU}(2), \mathrm{SU}(2))(\mathrm{U}(1), \mathrm{U}(1))$
  & $(\mathbf{2}, \mathbf{2}) (0, 2)$,   
    $(\mathbf{2}, \mathbf{2}) (0, -2)$,   
  & $(\mathbf{3}, \mathbf{1}) (-1, 0)$,   
    $(\mathbf{1}, \mathbf{3}) (-1, 0)$,   
    $(\mathbf{2}, \mathbf{2}) (1, -2)$,   
  \\ 
  & 
  & $(\mathbf{1}, \mathbf{1}) (2, 0)$,   
    $(\mathbf{1}, \mathbf{1}) (-2, 0)$  
  & $(\mathbf{2}, \mathbf{2}) (1, 2)$,  
    $(\mathbf{1}, \mathbf{1}) (-1, 4)$,   
    $(\mathbf{1}, \mathbf{1}) (-1, -4)$,   
  \\ 
  \hline
  (24)
  & $\mathrm{SU}(3)(\mathrm{U}(1), \mathrm{U}(1))$
  & $\mathbf{3} (0, 0)$,   
    $\bar{\mathbf{3}} (0, 0)$,   
    $\mathbf{1} (2, 0)$,   
  & $\mathbf{3} (1, -1)$,   
    $\mathbf{3} (-1, 1)$,   
    $\bar{\mathbf{3}} (1, 1)$,   
    $\bar{\mathbf{3}} (-1, -1)$,   
  \\ 
  & 
  & $\mathbf{1} (-2, 0)$,   
    $\mathbf{1} (0, 2)$,   
    $\mathbf{1} (0, -2)$  
  & $\mathbf{1} (1, -1)$,   
    $\mathbf{1} (-1, 1)$,  
    $\mathbf{1} (1, 1)$,   
    $\mathbf{1} (-1, -1)$   
  \\ 
  \hline
  (25)
  & $\mathrm{SU}(2)(\mathrm{U}(1), \mathrm{U}(1))$
  & $\mathbf{2} (-1, 2)$,   
    $\mathbf{2} (1, 2)$,   
    $\mathbf{2} (-1, -2)$,   
  & $\mathbf{3} (-1, 0)$,   
    $\mathbf{2} (2, 2)$,   
    $\mathbf{2} (0, 2)$,   
    $\mathbf{2} (0, -2)$,   
    $\mathbf{2} (2, -2)$,   
  \\ 
  & 
  & $\mathbf{2} (1, -2)$,   
    $\mathbf{1} (2, 0)$,   
    $\mathbf{1} (-2, 0)$  
  & $\mathbf{1} (-1, 4)$,  
    $\mathbf{1} (-1, -4)$,  
    $\mathbf{1} (-3, 0)$,  
    $\mathbf{1} (1, 0)$,  
    $\mathbf{1} (-1, 0)$  
  \\ 
  \hline
(26)
& $(\mathrm{SU}(2),\mathrm{SU}(2))(\mathrm{U}(1),\mathrm{U}(1),\mathrm{U}(1)) $ 
& $(\mathbf{2},\mathbf{2})(0,0,0)$, $(\mathbf{1},\mathbf{1})(2,0,0)$, 
& $(\mathbf{2},\mathbf{1})(1,1,1)$, $(\mathbf{2},\mathbf{1})(-1,-1,1)$, 
\\
&
& $(\mathbf{1},\mathbf{1})(-2,0,0)$, $(\mathbf{1},\mathbf{1})(0,2,0)$, 
& $(\mathbf{2},\mathbf{1})(1,-1,-1)$, $(\mathbf{2},\mathbf{1})(-1,1,-1)$, 
\\
&
& $(\mathbf{1},\mathbf{1})(0,-2,0)$, $(\mathbf{1},\mathbf{1})(0,0,2)$, 
& $(\mathbf{1},\mathbf{2})(1,-1,1)$, $(\mathbf{1},\mathbf{2})(-1,1,1)$, 
\\
&
& $(\mathbf{1},\mathbf{1})(0,0,-2)$ 
& $(\mathbf{1},\mathbf{2})(1,1,-1)$, $(\mathbf{1},\mathbf{2})(-1,-1,-1)$ 
\\ \hline 
(27)
& $(\mathrm{SU}(2), \mathrm{SU}(2))(\mathrm{U}(1), \mathrm{U}(1), \mathrm{U}(1)) $
& $(\mathbf{2},\mathbf{1})(3,0,0)$, $(\mathbf{2},\mathbf{1})(-3,0,0)$, 
& $(\mathbf{2},\mathbf{2})(0,0,-1)$, $(\mathbf{2},\mathbf{1})(0,3,1)$, 
\\
&
& $(\mathbf{1},\mathbf{2})(0,3,0)$, $(\mathbf{1},\mathbf{2})(0,-3,0)$, 
& $(\mathbf{2},\mathbf{1})(0,-3,1)$, $(\mathbf{1},\mathbf{2})(3,0,1)$, 
\\
&
& $(\mathbf{1},\mathbf{1})(0,0,2)$, $(\mathbf{1},\mathbf{1})(0,0,-2)$ 
& $(\mathbf{1},\mathbf{2})(-3,0,1)$, $(\mathbf{1},\mathbf{1})(3,3,-1)$, 
\\
&
&
& $(\mathbf{1},\mathbf{1})(-3,3,-1)$, $(\mathbf{1},\mathbf{1})(3,-3,-1)$,  
\\ 
&
&
& $(\mathbf{1},\mathbf{1})(-3,-3,-1)$
\\ \hline 
(28) 
&  $\mathrm{SU}(3) (\mathrm{U}(1), \mathrm{U}(1), \mathrm{U}(1) )$ 
& $\mathbf{3}(-4,0,0)$, $\bar{\mathbf{3}}(4,0,0)$,
& $\mathbf{3}(2,-1,1)$, $\mathbf{3}(2,1,-1)$,
\\
&
& $\mathbf{1}(0,2,0)$, $\mathbf{1}(0,-2,0)$,
& $\bar{\mathbf{3}}(-2,1,1)$, $\bar{\mathbf{3}}(-2,-1,-1)$,
\\
&
& $\mathbf{1}(0,0,2)$, $\mathbf{1}(0,0,-2)$
& $\mathbf{1}(6,1,1)$, $\mathbf{1}(-6,-1,1)$,
\\
&
&
& $\mathbf{1}(-6,1,-1)$, $\mathbf{1}(6,-1,-1)$
\\ \hline 
(29) 
& $\mathrm{SU}(2)( \mathrm{U}(1), \mathrm{U}(1), \mathrm{U}(1))$ 
& $\mathbf{3}(2,0,0)$, $\mathbf{3}(-2,0,0)$,
& $\mathbf{3}(-1,1,1)$, $\mathbf{3}(-1,-1,-1)$,
\\
&
& $\mathbf{1}(0,2,0)$, $\mathbf{1}(0,-2,0)$,
& $\mathbf{3}(1,1,-1)$, $\mathbf{3}(1,-1,1)$,
\\
&
& $\mathbf{1}(0,0,2)$, $\mathbf{1}(0,0,-2)$
& $\mathbf{1}(3,1,1)$, $\mathbf{1}(3,-1,-1)$,
\\
&
&
& $\mathbf{1}(-3,1,-1)$, $\mathbf{1}(-3,-1,1)$
\\ \hline
(30)
& $\mathrm{SU}(2)( \mathrm{U}(1), \mathrm{U}(1), \mathrm{U}(1))$
& $\mathbf{2}(1,0,0)$, $\mathbf{2}(-1,0,0)$,
& $\mathbf{2}(1,1,-1)$, $\mathbf{2}(1,-1,1)$,
\\
&
& $\mathbf{1}(2,0,0)$, $\mathbf{1}(-2,0,0)$,
& $\mathbf{2}(-1,1,1)$, $\mathbf{2}(-1,-1,-1)$,
\\
&
& $\mathbf{1}(0,2,0)$, $\mathbf{1}(0,-2,0)$,
& $\mathbf{1}(2,1,1)$, $\mathbf{1}(2,-1,-1)$,
\\
&
& $\mathbf{1}(0,0,2)$, $\mathbf{1}(0,0,-2)$
& $\mathbf{1}(-2,1,-1)$, $\mathbf{1}(-2,-1,1)$,
\\
&
&
& $\mathbf{1}(0,1,1)$, $\mathbf{1}(0,-1,-1)$,
\\
&
&
& $\mathbf{1}(0,1,-1)$, $\mathbf{1}(0,-1,1)$
\\ \hline 
(31)
& $\mathrm{SU}(2)(\mathrm{U}(1),\mathrm{U}(1),\mathrm{U}(1))$
& $\mathbf{2}(3,0,0)$, $\mathbf{2}(-3,0,0)$,
& $\mathbf{2}(0,1,3)$, $\mathbf{2}(0,1,-3)$,
\\
&
& $\mathbf{1}(0,2,0)$, $\mathbf{1}(0,-2,0)$,
& $\mathbf{2}(0,0,0)$, $\mathbf{2}(0,-2,0)$,
\\
&
& $\mathbf{1}(0,1,3)$, $\mathbf{1}(0,-1,-3)$,
& $\mathbf{1}(3,2,0)$, $\mathbf{1}(-3,2,0)$,
\\
&
& $\mathbf{1}(0,1,-3)$, $\mathbf{1}(0,-1,3)$
& $\mathbf{1}(3,0,0)$, $\mathbf{1}(-3,0,0)$,
\\
&
&
& $\mathbf{1}(3,-1,3)$, $\mathbf{1}(-3,-1,3)$,
\\
&
&
& $\mathbf{1}(3,-1,-3)$, $\mathbf{1}(-3,-1,-3)$
\\ \hline 
(32) 
& $(\mathrm{U}(1),\mathrm{U}(1),\mathrm{U}(1))$
& $(2,0,0)$, $(-2,0,0)$, $(0,-2,0)$,
& $(3,1,-1)$, $(3,-1,1)$, $(-3,-1,-1)$,
\\ 
&
& $(0,2,0)$, $(0,0,2)$, $(0,0,-2)$,
& $(-3,1,1)$, $(1,3,1)$, $(-1,-3,1)$,
\\
&
& $(2,2,0)$, $(-2,-2,0)$, $(2,-2,0)$,
& $(-1,3,-1)$, $(1,-3,1)$, $(-1,1,1)$,
\\
&
& $(-2,2,0)$
& $(1,-1,1)$, $(1,-1,-1)$, $(-1,1,1)$,
\\
&
&  
& $(1,1,1)$, $(-1,-1,1)$, $(1,1,-1)$,
\\
&
&
& $(-1,-1,-1)$ 
\\ \hline
\end{tabular}
\end{center}
\end{table}
\begin{table}
  \addtocounter{table}{-1}
  \caption{(Continued.)}
\begin{center}
 \let\tabularsize\scriptsize
  \renewcommand{\arraystretch}{1.2}
\begin{tabular}{llll}
  \hline
  $S/R$
  &
  & Branches of $\mathbf{10}$
  & Branches of $\mathbf{16}$
  \\ \hline \hline
(33)
& $\mathrm{SU}(2)(\mathrm{U}(1),\mathrm{U}(1),\mathrm{U}(1),\mathrm{U}(1))$
& $\mathbf{2}(3,0,0,0)$, $\mathbf{2}(-3,0,0,0)$,
& $\mathbf{2}(0,1,-1,1)$, $\mathbf{2}(0,-1,1,1)$,
\\
&
& $\mathbf{1}(0,2,0,0)$, $\mathbf{1}(0,-2,0,0)$,
& $\mathbf{2}(0,1,1,-1)$, $\mathbf{2}(0,-1,-1,-1)$,
\\
&
& $\mathbf{1}(0,0,2,0)$, $\mathbf{1}(0,0,-2,0)$,
& $\mathbf{1}(3,1,1,1)$, $\mathbf{1}(-3,1,1,1)$,
\\
&
& $\mathbf{1}(0,0,0,2)$, $\mathbf{1}(0,0,0,-2)$
& $\mathbf{1}(3,-1-1,1)$, $\mathbf{1}(-3,-1,-1,1)$,
\\
&
&
& $\mathbf{1}(3,1,-1,-1)$, $\mathbf{1}(-3,1,-1,-1)$,
\\
&
&
& $\mathbf{1}(3,-1,1,-1)$, $\mathbf{1}(-3,-1,1,-1)$
\\ \hline
(34)
& $(\mathrm{U}(1),\mathrm{U}(1),\mathrm{U}(1),\mathrm{U}(1),\mathrm{U}(1))$
& $(2,0,0,0,0)$, $(-2,0,0,0,0)$,
& $(1,1,1,-1,1)$, $(-1,-1,1,-1,1)$,
\\
&
& $(0,2,0,0,0)$, $(0,-2,0,0,0)$,
& $(1,1,-1,1,1)$, $(-1,-1,-1,1,1)$,
\\
&
& $(0,0,2,0,0)$, $(0,0,-2,0,0)$,
& $(1,1,1,1,-1)$, $(-1,-1,1,1,-1)$,
\\
&
& $(0,0,0,2,0)$, $(0,0,0,-2,0)$,
& $(1,1,-1,-1,-1)$, $(-1,-1,-1,-1,-1)$,
\\
&
& $(0,0,0,0,2)$, $(0,0,0,0,-2)$
& $(1,-1,1,1,1,)$, $(-1,1,1,1,1)$,
\\
&
&
& $(1,-1,-1,-1,1)$, $(-1,1,-1,-1,1)$,
\\
&
&
& $(1,-1,1,-1,-1)$, $(-1,1,1,-1,-1)$,
\\
&
&
& $(1,-1,-1,1,-1)$, $(-1,1,-1,1,-1)$
\\ \hline
(35)
& $(\mathrm{U}(1),\mathrm{U}(1),\mathrm{U}(1),\mathrm{U}(1))$
& $(1,3,0,0)$, $(-1,-3,0,0)$,
& $(2,0,-1,1)$, $(-2,0,1,1)$,
\\
&
& $(-1,3,0,0)$, $(1,-3,0,0)$,
& $(2,0,1,-1)$, $(-2,0,-1,-1)$,
\\
&
& $(2,0,0,0)$, $(-2,0,0,0)$,
& $(0,0,-1,1)$, $(0,0,1,1)$,
\\
&
& $(0,0,2,0)$, $(0,0,-2,0)$,
& $(0,0,1,-1)$, $(0,0,-1,-1)$,
\\
&
& $(0,0,0,2)$, $(0,0,0,-2)$
& $(1,3,1,1)$, $(1,-3,1,1)$,
\\
&
&
& $(-1,3,-1,1)$, $(-1,-3,-1,1)$,
\\
&
&
& $(1,3,-1,-1)$, $(1,-3,-1,-1)$,
\\
&
&
& $(-1,3,1,-1)$, $(-1,-3,1,-1)$
\\ \hline
\end{tabular}
\end{center}
\end{table}
\endgroup

The representation $F$ of $G$ for the fermions should be either
complex or pseudoreal but not real, since the fermions of real
representation do not allow the Majorana condition when $D = 14$ and
induces doubled fermion contents after the dimensional
reduction~\cite{Chapline:1982wy,Chiral-con:C14}.
Table \ref{representations} lists the candidate groups $G$
and their complex and pseudoreal representations.
Here we consider the dimensions of fermion representations less than 1025 
since even larger representations yield numerous higher dimensional representations 
of fermion, under the gauge group of the SM or GUTs, in the four-dimensions.
The representations in this table are the candidates of $F$.
\begin{table}
  \caption{
    The gauge groups that have either complex or pseudoreal
    representations and their complex and pseudoreal representations
    whose dimension is no larger than 1024 \cite{MckayPatera}.
    The groups $\mathrm{SU}(8)$ and $\mathrm{SU}(9)$ are not listed
    here since they do not lead to the four-dimensional gauge
    group of our interest for any of $S/R$ in
    Table~\ref{table_10dim_coset_spaces}.
  }
\label{representations}
\begin{center}
\begin{scriptsize}
\begin{tabular}{lll}
  \hline
  Group
  & Complex representations
  & Pseudoreal representations
  \\ \hline \hline
  $\mathrm{SU}(7)$
  & $\mathbf{21}$,
    $\mathbf{28}$,
    $\mathbf{35}$,
    $\mathbf{84}$,
    $\mathbf{112}$,
    $\mathbf{140}$,
    $\cdots$
  &
  \\ \hline
  $\mathrm{SO}(12)$
  &
  & $\mathbf{32}$,
    $\mathbf{32'}$,
    $\mathbf{352}$,
    $\mathbf{352'}$
  \\ \hline 
  $\mathrm{SO}(13)$
  &
  & $\mathbf{64}$, $\mathbf{768}$
  \\ \hline
  $\mathrm{Sp}(12)$
  &
  & $\mathbf{208}$, $\mathbf{364}$
  \\ \hline 
  $\mathrm{E}_{6}$
  & $\mathbf{27}$, $\mathbf{351}$, $\mathbf{351'}$
  &
  \\ \hline
  $\mathrm{SO}(14)$
  & $\mathbf{64}$, $\mathbf{832}$
  &
  \\ \hline 
  $\mathrm{Sp}(14)$
  &
  & $\mathbf{350}$, $\mathbf{560}$, $\mathbf{896}$
  \\ \hline
  $\mathrm{Sp}(16)$
  &
  & $\mathbf{544}$, $\mathbf{816}$
  \\ \hline
  $\mathrm{SU}(10)$
  & $\mathbf{45}$,
    $\mathbf{55}$,
    $\mathbf{120}$,
    $\mathbf{210}$,
    $\mathbf{220}$,
    $\mathbf{330}$,
    $\cdots$
  &
  \\ \hline 
  $\mathrm{SO}(18)$
  & $\mathbf{256}$
  &
  \\ \hline 
  $\mathrm{SO}(19)$
  &
  & $\mathbf{512}$
  \\ \hline 
  $\mathrm{Sp}(18)$
  &
  & $\mathbf{798}$
  \\ \hline
  $\mathrm{SO}(20)$
  &
  & $\mathbf{512}$
  \\ \hline 
  $\mathrm{SO}(21)$
  &
  & $\mathbf{1024}$
  \\ \hline
\end{tabular}
\end{scriptsize}
\end{center}
\end{table}

We constrain the gauge group $G$ by the following two criteria once we
choose $S/R$ out of the coset spaces listed in
Table~\ref{table_10dim_coset_spaces}.
First, $G$ should have an embedding of $R$ whose centralizer $C_G(R)$
is appropriate as a candidate of the four-dimensional gauge group $H$
(recall Eq.~(\ref{eq:H-is-centralizer})).
In this paper, we consider the following groups as candidates of $H$:
the GUT gauge groups such as $\mathrm{E}_{6}, \mathrm{SO}(10)$, and
$\mathrm{SU}(5)$;
the SM gauge group $\mathrm{SU}(3) \times \mathrm{SU}(2) \times
\mathrm{U}(1)$;
and those with an extra $\mathrm{U}(1)$.
Secondly, we consider only the regular subgroup of $G$ when we decompose
it to embed $R$.
We then find that no candidate of $G$ and $S/R$
that satisfy this requirement gives
$\mathrm{E}_{6}$, $\mathrm{E}_{6} \times \mathrm{U}(1)$, and
$\mathrm{SU}(5)$ as $H$.
We notice that the number of $\mathrm{U}(1)$'s in $R$ must be no
more than that in $H$, since the $\mathrm{U}(1)$'s in $R$ is also a
part of its centralizer, \textit{i.e.} a part of $H$.
We can thus exclude (26) -- (35) in Table~\ref{table_10dim_coset_spaces}.
The candidates of $G$ for each $S/R$ satisfying the above conditions are
summarized in Table~\ref{pairs}.
\begingroup
\begin{table}
  \caption{
    The allowed candidates of the gauge group $G$ for each choice of $H$
    and $S/R$.
    The top row indicates $H$ and the left column indicates $S/R$ by the
    number assigned in Table~\ref{table_10dim_coset_spaces}.
  }
 \label{pairs}
\begin{center}
\let\tabularsize\scriptsize
\begin{tabular}{l||l|l|l|l|l}
  \hline
  & $\mathrm{SO}(10)$
  & $\mathrm{SO}(10) \times \mathrm{U}(1)$
  & $\mathrm{SU}(5) \times \mathrm{U}(1)$
  & $\mathrm{SU}(3) \times \mathrm{SU}(2) \times \mathrm{U}(1)$
  & $\mathrm{SU}(3) \times \mathrm{SU}(2) \times \mathrm{U}(1) \times \mathrm{U}(1)$
  \\ \hline \hline
  (1)
  & $\mathrm{SO}(20)$
  &
  &
  &
  &
  \\ \hline 
  (2)
  & $\mathrm{SO}(20)$
  &
  &
  &
  &
  \\ \hline           
  (4)
  &
  & $\mathrm{SO}(20), \mathrm{SO}(21)$
  & $\mathrm{SU}(10)$
  &
  & $\mathrm{SU}(10), \mathrm{SO}(18), \mathrm{SO}(19)$
  \\ \hline  
  (5)
  &
  & $\mathrm{SO}(20), \mathrm{SO}(21)$
  & $\mathrm{SO}(18), \mathrm{SO}(19)$
  &
  & $\mathrm{SO}(18), \mathrm{SO}(19)$
  \\ \hline 
  (6)
  &
  & $\mathrm{SO}(20), \mathrm{SO}(21)$
  & $\mathrm{SO}(19)$
  &
  & $\mathrm{SO}(18), \mathrm{SO}(19), \mathrm{Sp}(18)$
  \\ \hline 
  (7)
  &
  &
  &
  &
  & $\mathrm{SO}(19), \mathrm{Sp}(18)$
  \\ \hline
  (8)
  &
  & $\mathrm{SO}(20), \mathrm{SO}(21)$
  & $\mathrm{SO}(18), \mathrm{SO}(19), \mathrm{Sp}(18)$
  & $\mathrm{Sp}(16)$
  & $\mathrm{SO}(18), \mathrm{SO}(19), \mathrm{Sp}(18)$
  \\ \hline
  (9)
  &
  &
  &
  &
  & $\mathrm{Sp}(16)$
  \\ \hline
  (10)
  &
  & $\mathrm{SO}(18), \mathrm{SO}(19)$
  & $\mathrm{Sp}(16)$
  & $\mathrm{SO}(14), \mathrm{Sp}(14)$
  & $\mathrm{Sp}(16)$
  \\ \hline
  (11)
  &
  & $\mathrm{SO}(18), \mathrm{SO}(19)$
  & $\mathrm{Sp}(16)$
  & $\mathrm{SO}(14), \mathrm{Sp}(14)$
  & $\mathrm{Sp}(16)$
  \\ \hline
  (12)
  &
  & $\mathrm{SO}(19)$
  & $\mathrm{Sp}(16)$
  & $\mathrm{Sp}(14)$
  & $\mathrm{Sp}(16)$
  \\ \hline
  (13)
  &
  &
  & $\mathrm{SO}(14), \mathrm{Sp}(14)$
  & $\mathrm{SO}(13), \mathrm{Sp}(12)$
  & $\mathrm{SO}(14), \mathrm{Sp}(14)$
  \\ \hline
  (14)
  &
  &
  & $\mathrm{Sp}(14)$
  & $\mathrm{Sp}(12)$
  & $\mathrm{Sp}(16)$
  \\ \hline
  (15)
  &
  & $\mathrm{SO}(14)$
  & $\mathrm{SU}(7), \mathrm{SO}(13), \mathrm{Sp}(12)$
  & $\mathrm{SO}(10), \mathrm{SO}(11), \mathrm{Sp}(10)$
  & $\mathrm{SU}(7), \mathrm{SO}(12), \mathrm{SO}(13), \mathrm{Sp}(12), \mathrm{E}_{6}$
  \\ \hline
  (16)
  &
  &
  &
  &
  & $\mathrm{Sp}(16)$
  \\ \hline          
  (17)
  &
  &
  &
  &
  & $\mathrm{Sp}(16)$
  \\ \hline
  (18)
  &
  &
  &
  &
  & $\mathrm{SU}(9), \mathrm{Sp}(16)$
  \\ \hline
  (19)
  &
  &
  &
  &
  & $\mathrm{SU}(9), \mathrm{Sp}(16)$
  \\ \hline
  (20)
  &
  &
  &
  &
  & $\mathrm{SO}(14), \mathrm{Sp}(14)$
  \\ \hline
  (21)
  &
  &
  &
  &
  & $\mathrm{SO}(14), \mathrm{Sp}(14)$
  \\ \hline
  (22)
  &
  &
  &
  &
  & $\mathrm{SO}(14), \mathrm{Sp}(14)$
  \\ \hline
  (23)
  &
  &
  &
  &
  & $\mathrm{SO}(14), \mathrm{Sp}(14)$
  \\ \hline
  (24)
  &
  &
  &
  &
  & $\mathrm{SU}(8), \mathrm{Sp}(14)$
  \\ \hline
  (25)
  &
  &
  &
  &
  & $\mathrm{SU}(7), \mathrm{SO}(12), \mathrm{SO}(13), \mathrm{Sp}(12), \mathrm{E}_{6}$
  \\ \hline      
\end{tabular}
\end{center}
\end{table}
\endgroup

Careful consideration is necessary when there are more than one
branch in decomposing $G$ to its regular subgroup $H \times R$, since
the different decomposition branches lead to different representations of $H$ and $R$.
Two cases deserve close attention.
The first is the decomposition of $\mathrm{SO}(2n + 1)$. 
It has
essentially two distinct branches of decomposition, one being
\begin{equation}
  \mathrm{SO}(2n + 1)
  \supset \mathrm{SO}(2k_{0} + 1)
  \times \prod_{i} \mathrm{SO}(2k_{i}).
\label{eq:SOodd-via-SOodd}
\end{equation}
and the other being
\begin{equation}
  \mathrm{SO}(2n + 1)
  \supset \mathrm{SO}(2n)
  \supset \prod_{i} \mathrm{SO}(2k_{i}),
\label{eq:SOodd-via-SOeven}
\end{equation}
An example is the decomposition of $\mathrm{Sp}(4) \simeq
\mathrm{SO}(5)$ into $\mathrm{SU}(2) \times \mathrm{U}(1)$.
One of the two branches of decomposition is
\begin{math}
  \mathrm{Sp}(4) \supset \mathrm{SU}(2) \times \mathrm{U}(1),
\end{math}
which is equivalent to
\begin{math}
  \mathrm{SO}(5) \supset \mathrm{SO}(3) \times \mathrm{SO}(2),
\end{math}
corresponding to
Eq.~(\ref{eq:SOodd-via-SOodd}).
The other branch is
\begin{math}
  \mathrm{Sp}(4) 
  \simeq \mathrm{SO}(5)
  \supset \mathrm{SO}(4) 
  \simeq \mathrm{SU}(2) \times \mathrm{SU}(2)
  \supset \mathrm{SU}(2) \times \mathrm{U}(1),
\end{math}
corresponding to Eq.~(\ref{eq:SOodd-via-SOeven}).
The two branches of decomposition lead to different branching of the
representations.
The second is the normalization of U(1) charge. 
The different normalizations provide different 
representations of $H$ for four-dimensional fields.

\subsection{$H=\mathrm{SO}(10) ( \times \mathrm{U}(1))$}  

First we search for viable $\mathrm{SO}(10)$ models in four
dimensions.
We list below the combinations of $S/R$, $G$ and $F$ that provide
$H = \mathrm{SO}(10)(\times \mathrm{U}(1))$ and the representations which
contain field contents of the SM for the scalars and the fermions.
We indicate the coset $S/R$ with its number assigned in Table~\ref{table_10dim_coset_spaces}
The embedding of $R$ into $G$ is shown for each candidates since this
embedding uniquely determines all the representations of the scalars
and fermions in the four-dimensional theory.
In Table~\ref{SO(10)-result}, we show all the field contents in four
dimensions for each combination of $(S/R, G, F)$.

\begin{table} 
  \caption{The field contents in four dimensions with
    $H = \mathrm{SO}(10) (\times \mathrm{U}(1))$ 
    for each combination of $(S/R, G, F)$.
    Coset spaces are indicated by the number assigned in
    Table~\ref{table_10dim_coset_spaces}.
    Numbers in a superscript of the representations denote its
    multiplicity.}
 \label{SO(10)-result}
\begin{center}
\let\tabularsize\footnotesize
\renewcommand{\arraystretch}{1.1}
\begin{tabular}{l|l|l||l|l|l} \hline 
  \multicolumn{3}{l||}{14D model} &
  \multicolumn{3}{l}{4D model}
  \\ \hline
  $S/R$
  & $G$
  & $F$
  & $H$
  & Scalars
  & Fermions
  \\ \hline \hline
  (1)
  & $\mathrm{SO}(20)$
  & $\mathbf{512}$
  & $\mathrm{SO}(10)$
  & $\mathbf{10}$
  & $\mathbf{16}$
  \\ \hline
  (2)
  & $\mathrm{SO}(20)$
  & $\mathbf{512}$
  & $\mathrm{SO}(10)$
  & $ \{ \mathbf{10} \}^{2}$
  & $ \{ \mathbf{16} \}^{2}$
  \\ \hline 
  (4)
  & $\mathrm{SO}(20)$
  & $\mathbf{512}$
  & $\mathrm{SO}(10) \times \mathrm{U}(1)$
  & $\mathbf{10}(2)$, $\mathbf{10}(-2)$
  & $\mathbf{16}(-1)$, $\mathbf{16}(3)$, $\mathbf{16}(-5)$
  \\ \hline 
  (5)
  & $\mathrm{SO}(20)$
  & $\mathbf{512}$
  & $\mathrm{SO}(10) \times \mathrm{U}(1)$
  & $\mathbf{10}(0)$, $\mathbf{10}(2)$, $\mathbf{10}(-2)$
  & $\mathbf{16}(1)$, $\mathbf{16}(-1)$
  \\ \hline
  (6)
  & $\mathrm{SO}(20)$
  & $\mathbf{512}$
  & $\mathrm{SO}(10) \times \mathrm{U}(1)$
  & $\mathbf{10}(0)$, $\mathbf{10}(1)$, $\mathbf{10}(-1)$
  & $\mathbf{16}(0)$, $\mathbf{16}(1)$, $\mathbf{16}(-1)$
  \\ \hline
  (8)
  & $\mathrm{SO}(20)$
  & $\mathbf{512}$
  & $\mathrm{SO}(10) \times \mathrm{U}(1)$
  & $\mathbf{10}(0)$, $\mathbf{10}(0)$, $\mathbf{10}(2)$, $\mathbf{10}(-2)$
  & $\mathbf{16}(1)$, $\mathbf{16}(1)$, $\mathbf{16}(-1)$, $\mathbf{16}(-1)$
  \\ \hline
  (10)
  & $\mathrm{SO}(18)$
  & $\mathbf{256}$
  & $\mathrm{SO}(10) \times \mathrm{U}(1)$
  & $\mathbf{10}(0)$
  & $\mathbf{16}(3)$, $\mathbf{16}(-3)$,
    $\overline{\mathbf{16}}(-3)$, $\overline{\mathbf{16}}(3)$
  \\ \hline
  (11)
  & $\mathrm{SO}(18)$
  & $\mathbf{256}$
  & $\mathrm{SO}(10) \times \mathrm{U}(1)$
  & $\mathbf{10}(0)$
  & $\mathbf{16}(2)$, $\mathbf{16}(-2)$,
    $\overline{\mathbf{16}}(-2)$, $\overline{\mathbf{16}}(2)$
  \\ \hline 
  (11)
  & $\mathrm{SO}(19)$
  & $\mathbf{512}$
  & $\mathrm{SO}(10) \times \mathrm{U}(1)$
  & $\mathbf{10}(0)$, $\mathbf{10}(2)$, $\mathbf{10}(-2)$
  & $\mathbf{16}(1)$, $\mathbf{16}(-1)$,
    $\overline{\mathbf{16}}(1)$, $\overline{\mathbf{16}}(-1)$
  \\ \hline
  (15)
  & $\mathrm{SO}(14)$
  & $\mathbf{64}$
  & $\mathrm{SO}(10) \times \mathrm{U}(1)$
  & (a): $\mathbf{10}(1)$, $\mathbf{10}(-1)$, $\mathbf{1}(2)$, $\mathbf{1}(-2)$
  & (a): $\mathbf{16}(0)$, $\mathbf{16}(1)$, $\mathbf{16}(-1)$,
  \\ 
  &
  &
  &
  &
  & \hspace{1.7em} $\overline{\mathbf{16}}(0)$,
    $\overline{\mathbf{16}}(-1)$, $\overline{\mathbf{16}}(1)$
  \\ 
  &
  &
  &
  & (b): $\mathbf{10}(3)$, $\mathbf{10}(-3)$
  & (b): $\mathbf{16}(0)$, $\mathbf{16}(3)$, $\mathbf{16}(-3)$,
  \\ 
  &
  &
  &
  &
  & \hspace{1.7em} $\overline{\mathbf{16}}(0)$,
    $\overline{\mathbf{16}}(-3)$, $\overline{\mathbf{16}}(3)$
  \\ \hline
\end{tabular}
\end{center}
\end{table}

%
\vspace{2mm} \noindent %
(a) $S/R \, \textrm{(11)} = \mathrm{Sp}(4) \times \mathrm{Sp}(4) / [\mathrm{SU}(2)
\times \mathrm{U}(1)]_{\textrm{non-max}} \times [ \mathrm{SU}(2)
\times \mathrm{SU}(2)]$, $G = \mathrm{SO}(19)$, and $F =
\mathbf{512}$.

We embed $R$ in the subgroup $\mathrm{SU}(2) \times \mathrm{SU}(2)
\times \mathrm{SU}(2) \times \mathrm{U}(1)$ of $\mathrm{SO}(19)$
according to the decomposition
\begin{eqnarray}
  \mathrm{SO}(19)
  &\supset& \mathrm{SO}(10) \times \mathrm{SO}(9) \nonumber
  \\ & \supset & \mathrm{SO}(10) \times \mathrm{SU}(4) \times \mathrm{SU}(2) \nonumber 
  \\ & \supset & \mathrm{SO}(10) \times \mathrm{SU}(2)
            \times \mathrm{SU}(2) \times \mathrm{SU}(2)
            \times \mathrm{U}(1).
\label{decomposition_branch_1} 
\end{eqnarray}

Notice that there is another branch of the decomposition such as 
\begin{eqnarray}
  \mathrm{SO}(19)
  &\supset& \mathrm{SO}(18)
  \supset \mathrm{SO}(10) \times \mathrm{SO}(8) \nonumber 
  \\ 
  & \supset & \mathrm{SO}(10) \times \mathrm{SU}(2) \times \mathrm{SU}(2)
            \times \mathrm{SU}(2) \times \mathrm{SU}(2) \nonumber 
  \\ 
  &\supset & \mathrm{SO}(10) \times \mathrm{SU}(2) \times \mathrm{SU}(2)
          \times \mathrm{SU}(2) \times \mathrm{U}(1).
\label{decomposition_branch_2}
\end{eqnarray}
As mentioned at the beginning of this section, it gives different
representations of the subgroup $\mathrm{SO}(10) \times \mathrm{SU}(2)
\times \mathrm{SU}(2) \times \mathrm{SU}(2) \times \mathrm{U}(1)$ for
a representation of $\mathrm{SO}(19)$.
For example, the adjoint representation $\mathbf{171}$ of
$\mathrm{SO}(19)$ is decomposed according to decomposition branch
Eq.~(\ref{decomposition_branch_1}) and
Eq.~(\ref{decomposition_branch_2}) as follows \cite{MckayPatera,Slansky:1981yr}:

\begin{eqnarray}
  \mathbf{171}
  &=&
  (\mathbf{45}, \mathbf{1}, \mathbf{1}, \mathbf{1})(0)
  + (\mathbf{1}, \mathbf{3}, \mathbf{1}, \mathbf{1})(0) \nonumber
  \\ & & \;
  + (\mathbf{1}, \mathbf{1}, \mathbf{3}, \mathbf{1})(0)
  + (\mathbf{1}, \mathbf{1}, \mathbf{1}, \mathbf{3})(0)
  + (\mathbf{1}, \mathbf{1}, \mathbf{1}, \mathbf{1})(0) \nonumber
  \\ & & \;
  + (\mathbf{1}, \mathbf{2}, \mathbf{2}, \mathbf{1})(2)
  + (\mathbf{1}, \mathbf{2}, \mathbf{2}, \mathbf{1})(-2)
  + (\mathbf{1}, \mathbf{2}, \mathbf{2}, \mathbf{3})(0) \nonumber
  \\ & & \;
  + (\mathbf{1}, \mathbf{1}, \mathbf{1}, \mathbf{3})(2)
  + (\mathbf{1}, \mathbf{1}, \mathbf{1}, \mathbf{3})(-2) \nonumber
  \\ & & \;
  + (\mathbf{10}, \mathbf{2}, \mathbf{2}, \mathbf{1})(0)
  + (\mathbf{10}, \mathbf{1}, \mathbf{1}, \mathbf{1})(2) \nonumber
  \\ & & \;
  + (\mathbf{10}, \mathbf{1}, \mathbf{1}, \mathbf{1})(-2)
  + (\mathbf{10}, \mathbf{1}, \mathbf{1}, \mathbf{3})(0),
\end{eqnarray}
\begin{eqnarray}
   \mathbf{171} &=&
  (\mathbf{45}, \mathbf{1}, \mathbf{1}, \mathbf{1})(0)
  + (\mathbf{1}, \mathbf{3}, \mathbf{1}, \mathbf{1})(0) \nonumber 
  \\ & & \;
  + (\mathbf{1}, \mathbf{1}, \mathbf{3}, \mathbf{1})(0)
  + (\mathbf{1}, \mathbf{1}, \mathbf{1}, \mathbf{3})(0)
  + (\mathbf{1}, \mathbf{1}, \mathbf{1}, \mathbf{1})(0) \nonumber
  \\ & & \;
  + (\mathbf{10}, \mathbf{1}, \mathbf{1}, \mathbf{1})(0)
  + (\mathbf{10}, \mathbf{2}, \mathbf{2}, \mathbf{1})(0) \nonumber 
  \\ & & \;
  + (\mathbf{1}, \mathbf{2}, \mathbf{2}, \mathbf{1})(0)
  + (\mathbf{1}, \mathbf{1}, \mathbf{1}, \mathbf{1})(2)
  + (\mathbf{1}, \mathbf{1}, \mathbf{1}, \mathbf{1})(-2) \nonumber
  \\ & & \;
  + (\mathbf{1}, \mathbf{2}, \mathbf{2}, \mathbf{2})(1)
  + (\mathbf{1}, \mathbf{2}, \mathbf{2}, \mathbf{2})(-1) \nonumber 
  \\ & & \;
  + (\mathbf{10}, \mathbf{1}, \mathbf{1}, \mathbf{2})(1)
  + (\mathbf{10}, \mathbf{1}, \mathbf{1}, \mathbf{2})(-1) \nonumber
  \\ & & \;
  + (\mathbf{1}, \mathbf{1}, \mathbf{1}, \mathbf{2})(1)
  + (\mathbf{1}, \mathbf{1}, \mathbf{1}, \mathbf{2})(-1).
\end{eqnarray}
The singlets of $\mathrm{SU}(2) \times \mathrm{SU}(2) \times \mathrm{SU}(2) \times
\mathrm{U}(1)$, which are $(\mathbf{45}, \mathbf{1}, \mathbf{1},
\mathbf{1})(0)$ and $(\mathbf{10}, \mathbf{1}, \mathbf{1},
\mathbf{1})(0)$, form an adjoint representation of $\mathrm{SO}(11)$ which is 
$(\mathbf{55}, \mathbf{1}, \mathbf{1}, \mathbf{1})(0)$. 
This indicates that the centralizer of
$\mathrm{SU}(2) \times \mathrm{SU}(2) \times \mathrm{SU}(2) \times
\mathrm{U}(1)$
is not $H = \mathrm{SO}(10) \times \mathrm{U}(1)$ but  
$\mathrm{SO}(11) \times \mathrm{U}(1)$, which is irrelevant to 
our purpose.

\vspace{2mm} \noindent %
(b) $S/R \, \textrm{(15a)} = \mathrm{G}_{2} / \mathrm{SU}(2) \times
\mathrm{U}(1)$, $G = \mathrm{SO}(14)$, and $F = \mathbf{64}$.

We embed $R$ in the subgroup $\mathrm{SU}(2) \times
\mathrm{U}(1)$ of $G=\mathrm{SO}(14)$ according to the decomposition
\begin{equation}
\begin{split}
  \mathrm{SO}(14)
  & \supset \mathrm{SO}(10) \times \mathrm{SU}(2) \times \mathrm{SU}(2) \\
  & \supset \mathrm{SO}(10) \times \mathrm{SU}(2) \times \mathrm{U}(1).
\end{split}
\label{decomposition_SO(14)}
\end{equation}
There are two branches of embedding which leads to the field contents
of the SM in this case, owing to the freedom of the normalization of
$\mathrm{U}(1)$ charges as mentioned in the beginning part of this
section.
For example, the adjoint representation of $\mathrm{SO}(14)$ can be
decomposed according to Eq.~(\ref{decomposition_SO(14)}) as \cite{MckayPatera,Slansky:1981yr}
\begin{equation}
\begin{split}
  \mathbf{91}
  = &
  (\mathbf{45}, \mathbf{1})(0)
  + (\mathbf{1}, \mathbf{3})(0)
  + (\mathbf{1}, \mathbf{1})(0)
  \\ &
  + (\mathbf{1}, \mathbf{1})(2x)
  + (\mathbf{1}, \mathbf{1})(-2x)
  \\ &
  + (\mathbf{10}, \mathbf{2})(x)
  + (\mathbf{10}, \mathbf{2})(-x),
\end{split}
\label{dec91}
\end{equation}
where $x$ is an arbitrary number reflecting the freedom of the
normalization.
The choice of $x = 1$ and $x = 3$ leads to the scalar contents (a) and
(b) of Table~\ref{SO(10)-result} respectively, as can be seen by
comparing the $\mathrm{U}(1)$ charges of Eq.~(\ref{dec91}) with those
in the row (15a) of Table~\ref{so10_vector_and_spinor_with_u1u1}.

\vspace{2mm} \noindent %
(c) $S/R \, \textrm{(1)} =\mathrm{SO}(11) / \mathrm{SO}(10)$, $G =
\mathrm{SO}(20)$, and $F = \mathbf{512}$.

We embed $R$ in the subgroup $\mathrm{SO}(10)$ of $G =
\mathrm{SO}(20)$ according to the
decomposition 
\begin{equation}
  \mathrm{SO}(20) \supset \mathrm{SO}(10) \times \mathrm{SO}(10).
\end{equation}

\vspace{2mm} \noindent %
(d) $S/R \, \textrm{(2)} = \mathrm{SO}(7) \times \mathrm{Sp}(4) /
\mathrm{SO}(6) \times [\mathrm{SU}(2) \times \mathrm{SU}(2)]$, $G =
\mathrm{SO}(20)$, and $F = \mathbf{512}$.
 
We embed $R$ in the subgroup
$\mathrm{SU}(4) \times \mathrm{SU}(2) \times \mathrm{SU}(2)$
of
$G =\mathrm{SO}(20)$
according to the decomposition
\begin{equation}
\begin{split}
  \mathrm{SO}(20)
  & \supset \mathrm{SO}(10) \times \mathrm{SO}(10) \\ 
  & \supset \mathrm{SO}(10) \times \mathrm{SU}(4)
            \times \mathrm{SU}(2) \times \mathrm{SU}(2).
\end{split}
\end{equation}

\vspace{2mm} \noindent %
(e) $S/R \, \textrm{(4)} = \mathrm{SU}(6)/\mathrm{SU}(5) \times
\mathrm{U}(1)$, $G = \mathrm{SO}(20)$, and $F = \mathbf{512}$.

We embed $R$ in the subgroup
$\mathrm{SU}(5) \times \mathrm{U}(1)$  
of
$G = \mathrm{SO}(20)$         
according to the decomposition
\begin{equation}
\begin{split}
 \mathrm{SO}(20)
 & \supset \mathrm{SO}(10) \times \mathrm{SO}(10) \\
 & \supset \mathrm{SO}(10) \times \mathrm{SU}(5) \times \mathrm{U}(1).
\end{split}
\end{equation}

\vspace{2mm} \noindent %
(f) $S/R \, \textrm{(5)} = \mathrm{SO}(9) \times \mathrm{SU}(2) /
\mathrm{SO}(8) \times \mathrm{U}(1)$, $G=\mathrm{SO}(20)$,
and $F = \mathbf{512}$.

We embed $R$ in the subgroup $\mathrm{SO}(8) \times \mathrm{U}(1)$ 
of $G = \mathrm{SO}(20)$ 
according to the decomposition
\begin{equation}
\begin{split}
  \mathrm{SO}(20)
  & \supset \mathrm{SO}(10) \times \mathrm{SO}(10) \\ 
  & \supset \mathrm{SO}(10) \times \mathrm{SO}(8) \times \mathrm{U}(1).
\end{split}
\end{equation}

\vspace{2mm} \noindent %
(g) $S/R \, \textrm{(6)} = \mathrm{SO}(7) \times \mathrm{SU}(3) /
\mathrm{SO}(6) \times [ \mathrm{SU}(2) \times \mathrm{U}(1)]$, $G =
\mathrm{SO}(20)$, and $F = \mathbf{512}$.

We embed $R$ in the subgroup $\mathrm{SU}(4) \times \mathrm{SU}(2)
\times \mathrm{U}(1)$ 
of $G = \mathrm{SO}(20)$ 
according to the decomposition
\begin{equation}
\begin{split}
 \mathrm{SO(20)}
 & \supset \mathrm{SO}(10) \times \mathrm{SO}(10) \\ 
 & \supset \mathrm{SO}(10) \times \mathrm{SU}(4) \times \mathrm{SU}(2)
           \times \mathrm{SU}(2) \\
 & \supset \mathrm{SO}(10) \times \mathrm{SU}(4)
           \times \mathrm{SU}(2) \times \mathrm{U}(1)
\end{split}
\end{equation}

\vspace{2mm} \noindent %
(h) $S/R \, \textrm{(8)} = \{ \mathrm{Sp}(4) \}^{2} \times
\mathrm{SU}(2) / [\mathrm{SU}(2) \times \mathrm{SU}(2)]^{2} \times
\mathrm{U}(1)$, $G = \mathrm{SO}(20)$, and $F = \mathbf{512}$.

We embed $R$ in the subgroup
$\mathrm{SU}(2) \times \mathrm{SU}(2) \times \mathrm{SU}(2) \times
\mathrm{SU}(2) \times \mathrm{U}(1)$ 
of $G = \mathrm{SO}(20)$ 
according to the decomposition
\begin{equation}
\begin{split}
  \mathrm{SO(20)}
  &
  \supset \mathrm{SO}(10) \times \mathrm{SO}(10)
  \\ &
  \supset \mathrm{SO}(10) \times \mathrm{SU}(4) \times \mathrm{SU}(2)
          \times \mathrm{SU}(2)
  \\ &
  \supset \mathrm{SO}(10) \times \mathrm{SU}(2)' \times \mathrm{SU}(2)'
  \\ & \quad
          \times \mathrm{SU}(2) \times \mathrm{SU}(2) \times \mathrm{U}(1).
\end{split}
\end{equation}

\vspace{2mm} \noindent %
(i) $S/R \, \textrm{(10)} = \mathrm{Sp}(4) \times \mathrm{Sp}(4) /
[\mathrm{SU}(2) \times \mathrm{U}(1)]_{\textrm{max}} \times [
\mathrm{SU}(2) \times \mathrm{SU}(2)]$, $G = \mathrm{SO}(18)$, and $F
= \mathbf{256}$.

We embed $R$ in the subgroup
$\mathrm{SU}(2) \times \mathrm{SU}(2) \times \mathrm{SU}(2) \times \mathrm{U}(1)$ 
of
$G = \mathrm{SO}(18)$ 
according to the decomposition
\begin{eqnarray}
  \mathrm{SO}(18)
  & \supset & \mathrm{SO}(10) \times \mathrm{SO}(8) \nonumber 
  \\ 
  &\supset & \mathrm{SO}(10) \times \mathrm{SU}(2) \times \mathrm{SU}(2)
          \times \mathrm{SU}(2) \times \mathrm{SU}(2) \nonumber 
  \\ 
  & \supset & \mathrm{SO}(10) \times \mathrm{SU}(2) \times \mathrm{SU}(2)
          \times \mathrm{SU}(2) \times \mathrm{U}(1).
\end{eqnarray}

\vspace{2mm} \noindent %
(j) $S/R \, \textrm{(11)} = \mathrm{Sp}(4) \times \mathrm{Sp}(4) /
[\mathrm{SU}(2) \times \mathrm{U}(1)]_{\textrm{non-max}} \times [
\mathrm{SU}(2) \times \mathrm{SU}(2)]$, $G = \mathrm{SO}(18)$ and $F =
\mathbf{256}$.

We embed $R$ in the subgroup
$\mathrm{SU}(2) \times \mathrm{SU}(2) \times \mathrm{SU}(2) \times \mathrm{U}(1)$
of
$G = \mathrm{SO}(18)$ 
according to the decomposition
\begin{eqnarray}
  \mathrm{SO}(18)
  &\supset & \mathrm{SO}(10) \times \mathrm{SO}(8) \nonumber 
  \\ 
  &\supset & \mathrm{SO}(10) \times \mathrm{SU}(2) \times \mathrm{SU}(2)
          \times \mathrm{SU}(2) \times \mathrm{SU}(2) \nonumber
  \\ 
  &\supset & \mathrm{SO}(10) \times \mathrm{SU}(2) \times \mathrm{SU}(2)
          \times \mathrm{SU}(2) \times \mathrm{U}(1).
\end{eqnarray}

We find ten candidates of $(S/R, G, F)$ which give at least one
fermion with representation $\mathbf{16}$ and scalar with
$\mathbf{10}$ in four dimensions.
Other combinations of $(S/R, G, F)$ are excluded since they do not
provide both a representation $\mathbf{16}$ for fermions and a
representation $\mathbf{10}$ for scalars.

In many cases we obtain several $\mathbf{16}$s for fermions. 
Particularly interesting candidates among them are 
($G = \mathrm{SO}(20)$, $S/R \, \textrm{(4)}$, $F = \mathbf{512}$) and
($G = \mathrm{SO}(20)$, $S/R \, \textrm{(6)}$, $F = \mathbf{512}$). 
They give three $\mathbf{16}$s corresponding to three generations of
fermions.
In such cases the extra $\mathrm{U}(1)$ symmetry can be interpreted as
a family symmetry.

We obtain the scalar field in the $\mathbf{10}$ representation of
$\mathrm{SO}(10)$ in all cases.
This scalar field contains the SM Higgs. 
Notice, however, that no scalar content belongs to $\mathbf{16},
\mathbf{45}, \mathbf{126}, \cdots$, which are necessary to break
$\mathrm{SO}(10)$ to the SM gauge group.
This is inevitable for $H = \mathrm{SO}(10) (\times
\mathrm{U}(1))$. 
The gauge group $G$ for $H = \mathrm{SO}(10) (\times \mathrm{U}(1))$
is $\mathrm{SO}(N)$, and $\mathrm{SO}(10)$ appears in the decomposition
\begin{equation}
  \mathrm{SO}(N)
  \supset \mathrm{SO}(10) \times \mathrm{SO}(N-10)
  \supset \cdots.
\end{equation}
Only $\mathbf{1}$ or $\mathbf{10}$ representations of $\mathrm{SO}(10)$
are obtained from the adjoint representation of $\mathrm{SO}(N)$ under
the above decomposition.
Thus no scalar can break $\mathrm{SO}(10)$ to the SM gauge group. 
Fortunately, we can construct a phenomenologically acceptable model
without these scalar contents by employing the topological symmetry
breaking mechanism, known as Hosotani mechanism or Wilson flux
breaking mechanism 
\cite{10dim-Model:K12,10dim-Model:D14,Hosotani:1983xw,Hosotani:1983vn,W-breaking:E8,W-breaking:J8,W-breaking:B8,W-breaking:B8-2,Zoupanos:1987wj}.
This mechanism requires extra-dimensional spaces to be non-simply 
connected.
Hence we have to consider
the non-simply connected coset spaces such as $(S/R)/T$ instead of the simply connected ones, %
where $T$ is a suitable discrete symmetry group.

\subsection{ $H=\mathrm{SU}(5) \times \mathrm{U}(1)$}  
Secondly, we search for viable $\mathrm{SU}(5) \times \mathrm{U}(1)$
models in four dimensions.
We list below the combinations of $S/R$, $G$ and $F$ which provides
$H=\mathrm{SU}(5) \times \mathrm{U}(1)$ and representations which
contain field contents of the SM for the scalars and the fermions.
The embedding of $R$ into $G$ is shown for each
candidates 
since this embedding uniquely determines all the representations of
the scalars and fermions in the four-dimensional theory.
In Table~\ref{SU(5)-result}, we show all the field contents in four
dimensions for each combination of $(S/R, G, F)$.

\begin{table}
\caption{
  The field contents in four dimensions with
  $H = \mathrm{SU}(5) \times \mathrm{U}(1)$  
  for each combination of $(S/R, G, F)$.
  Coset spaces are indicated by the number assigned in
  Table~\ref{table_10dim_coset_spaces}.
}
\label{SU(5)-result}
\begin{small}
\begin{center}
\begin{tabular}{l|l|l||l|l} \hline 
  \multicolumn{3}{l||}{14D model} &
  \multicolumn{2}{l}{4D model}
  \\ \hline
  $S/R$
  & $G$
  & $F$
  & Scalars
  & Fermions
  \\ \hline
  (11)
  & $\mathrm{Sp}(16)$
  & $\mathbf{544}$
  & $\mathbf{15}(2)$, $\overline{\mathbf{15}}(-2)$,
    $\mathbf{5}(1)$, $\overline{\mathbf{5}}(-1)$,
  & $\{ \mathbf{24}(0) \}^{2}$, $\mathbf{10}(2)$,
    $\overline{\mathbf{10}}(-2)$, $\mathbf{5}(1)$,
  \\
  &
  &
  & $\mathbf{1}(0)$
  & $\overline{\mathbf{5}}(-1)$, $\{ \mathbf{1}(0) \}^{4}$
  \\ \hline 
  (14)
  & $\mathrm{Sp}(14)$
  & $\mathbf{350}$
  & $\mathbf{15}(-2)$, $\overline{\mathbf{15}}(2)$, $\mathbf{5}(-1)$,
    $\overline{\mathbf{5}}(1)$
  & $\mathbf{45}(1)$, $\overline{\mathbf{45}}(-1)$, $\mathbf{24}(0)$,
    $\mathbf{10}(3)$,
  \\
  &
  &
  &
  & $\mathbf{10}(-2)$, $\overline{\mathbf{5}}(1)$,
    $\mathbf{5}(1)$, $\overline{\mathbf{5}}(-1)$
  \\ \hline
  (15)
  & $\mathrm{Sp}(12)$
  & $\mathbf{208}$
  & $\mathbf{15}(2)$, $\overline{\mathbf{15}}(-2)$, $\mathbf{5}(1)$,
    $\overline{\mathbf{5}}(-1)$
  & $\mathbf{45}(1)$, $\overline{\mathbf{45}}(-1)$, $\mathbf{24}(0)$,
    $\mathbf{10}(-3)$,
  \\
  &
  &
  &
  & $\overline{\mathbf{10}}(3)$, $\mathbf{10}(2)$,
    $\overline{\mathbf{10}}(-2)$, $\mathbf{5}(1)$,
    $\overline{\mathbf{5}}(-1)$
  \\ \hline
\end{tabular}
\end{center}
\end{small}
\end{table}

\vspace{2mm} \noindent %
(a)
$S/R \textrm{(15)}= \mathrm{G}_2/\mathrm{SU}(2) \times
\mathrm{U}(1)$, $G = \mathrm{Sp}(12)$ and $F = \mathbf{208}$.

We embed $R$ in the subgroup $\mathrm{SU}(2) \times \mathrm{U}(1)$ of
$G = \mathrm{Sp}(12)$ according to the decomposition
\begin{equation}
\begin{split}
  \mathrm{Sp(12)}
  & \supset \mathrm{Sp}(10) \times \mathrm{Sp}(2) \\
  & \supset \mathrm{SU}(5) \times \mathrm{SU}(2) \times \mathrm{U}(1).
\end{split}
\end{equation}

\vspace{2mm} \noindent 
(b)
$S/R \textrm{(14)}= \mathrm{Sp}(6)/\mathrm{Sp}(4) \times \mathrm{U}(1)$, 
$G = \mathrm{Sp}(14)$, and $F = \mathbf{350}$.
 
We embed $R$ in the subgroup $\mathrm{Sp}(4) \times \mathrm{U}(1)$ of
$G = \mathrm{Sp}(14)$ according to the decomposition
\begin{equation}
\begin{split}
  \mathrm{Sp}(14)
  & \supset \mathrm{Sp}(10) \times \mathrm{Sp}(4) \\
  & \supset \mathrm{SU}(5) \times \mathrm{Sp}(4) \times \mathrm{U}(1).
\end{split}
\end{equation}

\vspace{2mm} \noindent  
(c)
$S/R \textrm{(11)}= \mathrm{Sp}(4) \times
\mathrm{Sp}(4)/[\mathrm{SU}(2) \times
\mathrm{U}(1)]_{\textrm{non-max}} \times [\mathrm{SU}(2) \times
\mathrm{SU}(2)]$, $G = \mathrm{Sp}(16)$, and $F = \mathbf{544}$.
 
We embed $R$ in the subgroup $\mathrm{SU}(2)' \times \mathrm{SU}(2)'
\times \mathrm{SU}(2) \times \mathrm{U}(1)$ of $G = \mathrm{Sp}(16)$
according to the decomposition
\begin{eqnarray}
  \mathrm{Sp(16)}
  &\supset & \mathrm{Sp}(10) \times \mathrm{Sp}(6) \nonumber 
  \\ 
  &\supset & \mathrm{Sp}(10) \times \mathrm{Sp}(4) \times \mathrm{SU}(2) \nonumber 
  \\
  & \supset & \mathrm{Sp}(10) \times \mathrm{SU}(2)'
            \times \mathrm{SU}(2)' \times \mathrm{SU}(2) \nonumber 
  \\
  & \supset & \mathrm{SU}(5) \times \mathrm{SU}(2)' \times \mathrm{SU}(2)'
            \times \mathrm{SU}(2) \times \mathrm{U}(1).
\end{eqnarray}

We find three candidates of $(S/R, G, F)$ that give at least one pair
of fermions with representation $\mathbf{10}$ and $\bar{\mathbf{5}}$,
and a scalar with $\mathbf{5}$ representation in four dimensions.
Other combinations of $(S/R, G, F)$ are excluded since they do not
provide these representations for fermions and scalars.

We obtain the scalar field in $\mathbf{5}$ representation of
$\mathrm{SU}(5)$ for all cases.
This scalar field contains the SM Higgs.
Notice, however, that no scalar contents belongs to $\mathbf{24},
\cdots$, which are necessary to break $\mathrm{SU}(5)$ to the SM gauge
group.
The lack of such scalars is a general feature for $H = \mathrm{SU}(5)
\times \mathrm{U}(1)$.
The gauge groups $G$ for $H = \mathrm{SU}(5) \times \mathrm{U}(1) $
are $\mathrm{SU}(N)$, $\mathrm{SO}(N)$, and $\mathrm{Sp}(N)$.
These groups are decomposed into subgroups including $\mathrm{SU}(5)
\times \mathrm{U}(1)$, and their adjoint representations are
decomposed accordingly as well:
\begin{equation}
\begin{gathered}
  \mathrm{SU}(N)
  \supset
  \mathrm{SU}(5) \times \mathrm{SU}(N - 5) \times \mathrm{U}(1)
  \supset \cdots \hspace{20mm}
  \\
  \begin{split}
    \mathrm{adj} \, \mathrm{SU}(5)
    & =
    (\mathbf{24}, \mathbf{1})(0)
    + (\mathbf{1}, \mathrm{adj} \, \mathrm{SU}(N - 1))(0)
    + (1, 1)(0)
    \\ & \quad
    + (\mathbf{5}, \overline{N - 5})(a)
    + (\overline{\mathbf{5}}, N - 5)(-a)
    \\ &
    = \cdots 
  \end{split}
\end{gathered}
\end{equation}
\begin{equation}
  \begin{gathered}
  \begin{aligned}
    \mathrm{SO}(N)
    & \supset \mathrm{SO}(10) \times \mathrm{SO}(N-10)
    \\ &
    \supset \mathrm{SU}(5) \times \mathrm{SO}(N - 10) \times \mathrm{U}(1) 
    \supset \cdots \hspace{50mm}
  \end{aligned}
  \\
  \begin{aligned}
    \mathrm{adj} \, \mathrm{SO}(N)
    &
    = (\mathbf{45}, \mathbf{1})
    + (\mathbf{1}, \mathrm{adj} \, \mathrm{SO}(N - 10))
    \\ & \quad
    + (\mathbf{10}, \mathbf{1})
    + (\mathbf{1}, N - 10)
    \\ &
    = (\mathbf{24}, \mathbf{1})(0)
    + (\mathbf{1}, \mathrm{adj} \, \mathrm{SO}(N - 10))(0)
    + (\mathbf{1}, \mathbf{1})(0)
    \\ & \quad
    + (\mathbf{10}, \mathbf{1})(4)
    + (\overline{\mathbf{10}}, \mathbf{1})(-4)
    + (\mathbf{5}, \mathbf{1})(2)
    + (\overline{\mathbf{5}}, \mathbf{1})(-2)
    + (\mathbf{1}, N - 10)(0)
    \\ &
    = \cdots
  \end{aligned}
  \end{gathered}
\end{equation}
\begin{equation}
\begin{gathered}
\begin{aligned}
  \mathrm{Sp}(2N)
  & \supset \mathrm{Sp}(10) \times \mathrm{Sp}(2N - 10)
  \\ &
  \supset \mathrm{SU}(5) \times \mathrm{Sp}(2N - 1) \times \mathrm{U}(1)
  \\ &
  \supset \cdots \hspace{100mm}
\end{aligned}
\\ 
\begin{aligned}
  \mathrm{adj} \, \mathrm{Sp}(2N)
  &
  = (\mathbf{55}, \mathbf{1})
  + (\mathbf{1}, \mathrm{adj} \, \mathrm{Sp}(2N - 10))
  \\ & \quad
  + (\mathbf{10}, \mathbf{1})
  + (\mathbf{1}, 2N - 10)
  \\ &
  = (\mathbf{24}, \mathbf{1})(0)
  + (\mathbf{1}, \mathrm{adj} \, \mathrm{Sp}(2N - 10))(0)
  + (\mathbf{1}, \mathbf{1})(0)
  \\ & \quad
  + (\mathbf{15}, \mathbf{1})(2)
  + (\overline{\mathbf{15}}, \mathbf{1})(-2)
  + (\mathbf{5}, \mathbf{1})(1)
  + (\bar{\mathbf{5}}, \mathbf{1})(-1)
  + (\mathbf{1}, N)(0)
  \\ & \quad
  = \cdots.
\end{aligned}
\end{gathered}
\end{equation}
Only $\mathbf{1}$, $\mathbf{5}$, $\mathbf{10}$, or $\mathbf{15}$
representation of $\mathrm{SU}(5)$ is obtained from the adjoint
representations of $\mathrm{SU}(N)$, $\mathrm{SO}(N)$, and
$\mathrm{Sp}(N)$ under the above decompositions.
Then, no scalar can break $\mathrm{SU}(5)$ to the SM gauge group.
Therefore we should employ the flux breaking mechanism to break
$\mathrm{SU}(5)$ to the SM gauge group.

\subsection{$H=\mathrm{SU}(3) \times \mathrm{SU}(2) \times
  \mathrm{U}(1)$ }

We find no viable candidate for $H = \mathrm{SU}(3) \times
\mathrm{SU}(2) \times \mathrm{U}(1)$.
We exclude the coset spaces (16) -- (35) in
Table~\ref{table_10dim_coset_spaces}.
They have two or more factors of $\mathrm{U}(1)$ in $R$, and these
$\mathrm{U}(1)$'s become the part of $H = C_G(R) = \mathrm{SU}(3) \times
\mathrm{SU}(2) \times \mathrm{U}(1)$, which has only one
$\mathrm{U}(1)$.
The single $\mathrm{U}(1)$ factor in $R$ becomes
$\mathrm{U}(1)_{Y}$ of the SM gauge group, hence the
decomposition of the spinor representation $\mathbf{16}$ of
$\mathrm{SO}(10)$ to $R$ need to have $\mathrm{U}(1)$ charges whose
ratio is $1:2:(-3):(-4):6$.
Referring to Table~\ref{so10_vector_and_spinor_with_u1u1}, we find
that the coset spaces (4) -- (15) do not have such $\mathrm{U}(1)$
charge and thus are excluded.
The explicit analysis of the remaining coset spaces (1), (2) and (3)
shows that they do not induce the SM either.
%


\subsection{$H = \mathrm{SU}(3) \times \mathrm{SU}(2) \times
  \mathrm{U}(1) \times \mathrm{U}(1)$}

Finally, we search for viable $\mathrm{SU}(3) \times \mathrm{SU}(2)
\times \mathrm{U}(1) \times \mathrm{U}(1)$ models in four dimensions.
We list below the combinations of $S/R$, $G$, and $F$ which provide
$H = \mathrm{SU}(3) \times \mathrm{SU}(2) \times \mathrm{U}(1) \times
\mathrm{U}(1)$ and representations of the SM scalars and fermions.
Embedding of $R$ in $G$ is also shown for each candidates.
Note that we can take a linear combination of the two $\mathrm{U}(1)$'s.
The $\mathrm{U}(1)$ charges in the decomposition are first chosen to
facilitate the decomposition of the group $G$, then combined to embed
$R$ into $G$, and subsequently organized again to reproduce the
hypercharge of the SM.
We explicitly show these linear recombinations of $\mathrm{U}(1)$ for
each candidates.
In Table~\ref{table:SMGroup-decomp-Yalpha}, we show all the field
contents in four dimensions for each combination of $(S/R, G, F)$.

\begin{table}
\caption{ %
  The field contents in four dimensions with $H = \mathrm{SU}(3) \times
  \mathrm{SU}(2) \times \mathrm{U}(1)_R \times \mathrm{U}(1)_A$.
  Coset spaces are indicated by the number assigned in
  Table~\ref{table_10dim_coset_spaces}.
  Numbers in a superscript of the representations denote its
  multiplicity.
}
\label{table:SMGroup-decomp-RA}
\begin{center}
\let\tabularsize\scriptsize
\begin{tabular}{l|l|l||l|l} \hline 
  \multicolumn{3}{l||}{14D model}
  & \multicolumn{2}{l}{4D model}
  \\ \hline
  $S/R$
  & $G$
  & $F$
  & Scalars
  & Fermions
  \\ \hline \hline
  %
  %
  (15a)
  & $\mathrm{Sp}(12)$
  & $\mathbf{364}$
  & $(\mathbf{1}, \mathbf{2})(-2, 3)$,
    $(\mathbf{1}, \mathbf{2})(2, -3)$,
  & $(\mathbf{15}, \mathbf{1})(-1, 4)$,
    $(\overline{\mathbf{15}}, \mathbf{1})(1, -4)$,
    $(\mathbf{10}, \mathbf{1})(-3, -12)$,
  \\
  &
  &
  & $(\mathbf{3}, \mathbf{1})(-1, -4)$,
    $(\bar{\mathbf{3}}, \mathbf{1})(1, 4)$,
  & $(\overline{\mathbf{10}}, \mathbf{1})(3, 12)$,
    $(\mathbf{3}, \mathbf{1})(-1, -4)$,
    $\{ (\bar{\mathbf{3}}, \mathbf{1})(1, 4) \}^{3}$,
  \\
  &
  &
  & $(\mathbf{6}, \mathbf{1})(-2, -8)$,
    $(\bar{\mathbf{6}}, \mathbf{1})(2, 8)$
  & $(\mathbf{1}, \mathbf{3})(0, 0)$,
    $(\mathbf{1}, \mathbf{1})(-4, 6)$,
    $\{ (\mathbf{1}, \mathbf{1})(0, 0) \}^{2}$,
  \\
  &
  &
  &
  & $(\mathbf{1}, \mathbf{2})(-2, 3)$,
    $(\mathbf{1}, \mathbf{2})(2, -3)$,
    $(\mathbf{3}, \mathbf{3})(-1, -4)$,
  \\
  &
  &
  &
  & $(\bar{\mathbf{3}}, \mathbf{3})(1, 4)$,
    $(\bar{\mathbf{3}}, \mathbf{1})(5, -2)$,
    $(\mathbf{3}, \mathbf{1})(-1, -4)$,
  \\
  &
  &
  &
  & $(\mathbf{3}, \mathbf{1})(3, -10)$,
    $(\bar{\mathbf{3}}, \mathbf{1})(-3, 10)$,
    $(\mathbf{3}, \mathbf{2})(-3, -1)$,
  \\
  &
  &
  &
  & $(\bar{\mathbf{3}}, \mathbf{2})(3, 1)$,
    $(\mathbf{3}, \mathbf{2})(1, -7)$,
    $(\bar{\mathbf{3}}, \mathbf{2})(-1, 7)$,
  \\
  &
  &
  &
  & $(\mathbf{8}, \mathbf{1})(0, 0)$,
    $(\mathbf{6}, \mathbf{1})(2, -8)$,
    $(\bar{\mathbf{6}}, \mathbf{1})(-2, 8)$
  \\ \hline
  %
  %
  (9)
  & $\mathrm{Sp}(16)$
  & $\mathbf{544}$
  & $(\mathbf{1}, \mathbf{2})(1, 0)$,
    $(\mathbf{1}, \mathbf{2})(-1, 0)$
  & $(\mathbf{1}, \mathbf{1})(-2, 0)$,
    $(\mathbf{1}, \mathbf{2})(1, 0)$,
    $\{ (\mathbf{1}, \mathbf{1})(0, 0) \}^{2}$,
  \\
  &
  &
  &
  & $(\bar{\mathbf{3}}, \mathbf{1})(2, -1)$,
    $(\mathbf{3}, \mathbf{1})(2, 1)$,
    $(\bar{\mathbf{3}}, \mathbf{2})(-1, -1)$,
  \\
  &
  &
  &
  & $(\mathbf{3}, \mathbf{2})(-1, 1)$,
    $\{ (\mathbf{3}, \mathbf{1})(0, 1) \}^{3}$,
    $\{ (\bar{\mathbf{3}}, \mathbf{1})(0, -1) \}^{3}$,
  \\
  &
  &
  &
  & $(\mathbf{8}, \mathbf{1})(0, 0)$,
    $(\mathbf{6}, \mathbf{1})(0, -1)$,
    $(\bar{\mathbf{6}}, \mathbf{1})(0, 1)$
  \\ \hline
  %
  %
  (15a)
  & $\mathrm{SO}(13)$
  & $\mathbf{768}$
  & $(\mathbf{1}, \mathbf{2})(3, 3)$,
    $(\mathbf{1}, \mathbf{2})(-3, -3)$,
  & $(\mathbf{3}, \mathbf{3})(-2, -4)$,
    $(\bar{\mathbf{3}}, \mathbf{3})(2, 4)$,
    $(\mathbf{1}, \mathbf{3})(0, -6)$,
    $(\mathbf{1}, \mathbf{3})(0, 6)$,
  \\
  &
  &
  & $(\mathbf{3}, \mathbf{1})(-2, -6)$, 
    $(\bar{\mathbf{3}}, \mathbf{1})(2, 6)$
  & $(\mathbf{3}, \mathbf{2})(1, 3)$,
    $(\bar{\mathbf{3}}, \mathbf{1})(-4, -6)$,
    $(\mathbf{3}, \mathbf{1})(-2, 0)$,
    $(\bar{\mathbf{3}}, \mathbf{1})(2, 0)$,
  \\
  &
  &
  &
  & $(\mathbf{3}, \mathbf{2})(1, 3)$,
    $(\bar{\mathbf{3}}, \mathbf{2})(-1, -3)$,
    $(\bar{\mathbf{3}}, \mathbf{2})(5, 3)$,
    $(\mathbf{1}, \mathbf{1})(0, -6)$,
  \\
  &
  &
  &
  & $(\mathbf{1}, \mathbf{1})(0, 6)$,
    $(\mathbf{1}, \mathbf{2})(3, -3)$,
    $(\mathbf{1}, \mathbf{2})(-3, 3)$,
  \\
  &
  &
  &
  & $(\mathbf{1}, \mathbf{2})(-3, -9)$,
    $(\mathbf{1}, \mathbf{2})(3, 9)$,
    $(\mathbf{3}, \mathbf{2})(1, 3)$,
  \\
  &
  &
  &
  & $(\bar{\mathbf{3}}, \mathbf{2})(-1, -3)$,
    $(\mathbf{3}, \mathbf{1})(-2, 0)$,
    $(\bar{\mathbf{3}}, \mathbf{1})(2, 0)$,
  \\
  &
  &
  &
  & $(\mathbf{1}, \mathbf{1})(0, 6)$,
    $(\mathbf{1}, \mathbf{1})(0, -6)$,
    $(\mathbf{1}, \mathbf{2})(3, -3)$,
    $(\mathbf{1}, \mathbf{2})(-3, 3)$,
  \\
  &
  &
  &
  & $(\mathbf{3}, \mathbf{1})(-2, 0)$,
    $(\bar{\mathbf{3}}, \mathbf{1})(2, 0)$,
    $(\mathbf{3}, \mathbf{2})(1, 3)$,
    $(\bar{\mathbf{3}}, \mathbf{2})(-1, -3)$,
  \\
  &
  &
  &
  & $(\bar{\mathbf{3}}, \mathbf{1})(-4, 6)$,
    $(\mathbf{3}, \mathbf{2})(1, -9)$,
    $(\bar{\mathbf{3}}, \mathbf{2})(-1, 9)$,
    $(\mathbf{6}, \mathbf{1})(2, 0)$,
  \\
  &
  &
  &
  & $(\bar{\mathbf{6}}, 1)(-2, 0)$,
    $(\mathbf{6}, \mathbf{2})(-1, -3)$,
    $(\bar{6}, 2)(1, 3)$,
    $(\mathbf{8}, \mathbf{1})(2, 0)$,
  \\
  &
  &
  &
  & $(\mathbf{8}, \mathbf{1})(-2, 0)$,
    $(\mathbf{8}, \mathbf{2})(-1, -3)$,
    $(\mathbf{8}, \mathbf{2})(1, 3)$,
  \\
  &
  &
  &
  & $(\mathbf{3}, \mathbf{1})(-2, 0)$,
    $(\bar{\mathbf{3}}, \mathbf{1})(2, 0)$,
    $(\mathbf{3}, \mathbf{2})(1, 3)$,
    $(\bar{\mathbf{3}}, \mathbf{2})(-1, -3)$,
  \\
  &
  &
  &
  & $(\mathbf{1}, \mathbf{1})(0, -6)$,
    $(\mathbf{1}, \mathbf{1})(0, 6)$,
    $(\mathbf{1}, \mathbf{2})(3, -3)$,
    $(\mathbf{1}, \mathbf{2})(-3, 3)$
  \\ \hline
  %
  %
  (14)
  & $\mathrm{Sp}(14)$
  & $\mathbf{350}$
  & $(\mathbf{1}, \mathbf{2})(-1, -9/2)$,
  & $(\mathbf{6}, \mathbf{1})(3, -1)$,
    $(\mathbf{8}, \mathbf{1})(0, 0)$,
    $(\mathbf{1}, \mathbf{1})(-2, -9)$,
  \\
  &
  &
  & $(\mathbf{1}, \mathbf{2})(1, 9/2)$,
  & $\{ (\mathbf{1}, \mathbf{1})(0, 0) \}^{2}$,
    $(\mathbf{3}, \mathbf{1})(-1, 10)$,
    $(\bar{\mathbf{3}}, \mathbf{1})(1, -10)$,
  \\
  &
  &
  & $(\mathbf{3}, \mathbf{2})(-2, 11/2)$,
  & $\{ (\bar{\mathbf{3}}, \mathbf{1})(3, -1) \}^{2}$,
    $\{ (\mathbf{1}, \mathbf{2})(-1, -9/2) \}^{2}$,
  \\
  &
  &
  & $(\bar{\mathbf{3}}, \mathbf{2})(2, -11/2)$,
  & $\{ (\mathbf{1}, \mathbf{2})(1, 9/2) \}^{3}$,
    $(\mathbf{3}, \mathbf{2})(-2, 11/2)$,
    $(\mathbf{1}, \mathbf{3})(0, 0)$,
  \\
  &
  &
  & $(\mathbf{1}, \mathbf{3})(-2, -9)$, $(\mathbf{1}, \mathbf{3})(2, 9)$
  & $(\bar{\mathbf{3}}, \mathbf{3})(3, -1)$
  \\ \hline
\end{tabular}
\end{center}
\end{table}
\begin{table}
\caption{ %
  The field contents in four dimensions with $H = \mathrm{SU}(3) \times
  \mathrm{SU}(2) \times \mathrm{U}(1)_Y \times \mathrm{U}(1)_{\alpha}$.
  Coset spaces are indicated by the number assigned in
  Table~\ref{table_10dim_coset_spaces}.
  Numbers in superscript of the representations denote its
  multiplicity.
  The $\mathrm{U}(1)$ charges are rearranged from
  those of Table~\ref{table:SMGroup-decomp-RA} so that the charge of
  $\mathrm{U}(1)_{Y}$ is proportional to the hypercharge of the
  Standard Model.
}
\label{table:SMGroup-decomp-Yalpha}
\begin{center}
\let\tabularsize\scriptsize
\begin{tabular}{l|l|l||l|l|l|l} \hline 
  \multicolumn{3}{l||}{14D model}
  & \multicolumn{4}{l}{4D model}
  \\ \hline
  &
  &
  & \multicolumn{2}{l|}{Scalars}
  & \multicolumn{2}{l}{Fermions}
  \\ \cline{4-7}
  $S/R$
  & $G$
  & $F$
  & SM fields & Extra fields
  & SM fields & Extra fields
  \\ \hline \hline
  %
  (15a)
  & $\mathrm{Sp}(12)$
  & $\mathbf{364}$
  & $(\mathbf{1}, \mathbf{2})(3, -32)$,
  & $(\mathbf{3}, \mathbf{1})(-2, -27)$,
  & $(\mathbf{3}, \mathbf{2})(1, -59)$,
  & $(\mathbf{15}, 1)(34/11, -11)$,
  \\
  &
  &
  & $(\mathbf{1}, \mathbf{2})(-3, 32)$
  & $(\bar{\mathbf{3}}, \mathbf{1})(2, 27)$,
  & $(\bar{\mathbf{3}}, \mathbf{1})(2, 27)$
  & $(\overline{\mathbf{15}}, \mathbf{1})(-34/11, 11)$,
  \\
  &
  &
  &
  & $(\mathbf{6}, \mathbf{1})(-4, -54)$,
  & $(\bar{\mathbf{3}}, \mathbf{1})(-4, 91)$
  & $(\mathbf{10}, \mathbf{1})(-6, -81)$,
    $(\overline{\mathbf{10}}, \mathbf{1})(6, 81)$,
  \\
  &
  &
  &
  & $(\bar{\mathbf{6}}, \mathbf{1})(4, 54)$
  & $(\mathbf{1}, \mathbf{2})(-3, 32)$
  & $\{ (\mathbf{3}, \mathbf{1})(-2, -27) \}^{2}$,
    $(\mathbf{1}, \mathbf{3})(0, 0)$,
  \\
  &
  &
  &
  &
  & $(\mathbf{1}, \mathbf{1})(6, -64)$
  & $\{ (\mathbf{1}, \mathbf{1})(0, 0) \}^{2}$,
    $(\mathbf{1}, \mathbf{2})(3, -32)$,
  \\
  &
  &
  &
  &
  &
  & $(\mathbf{3}, \mathbf{3})(-2, -27)$,
    $(\bar{\mathbf{3}}, \mathbf{3})(2, 27)$,
  \\
  &
  &
  &
  &
  &
  & $(\mathbf{3}, \mathbf{1})(-8, 37)$,
    $(\bar{\mathbf{3}}, \mathbf{1})(8, -37)$,
  \\
  &
  &
  &
  &
  &
  &    $(\bar{\mathbf{3}}, \mathbf{2})(-1, 59)$,
    $(\mathbf{3}, \mathbf{2})(-5, 5)$,
  \\
  &
  &
  &
  &
  &
  & $(\bar{\mathbf{3}}, \mathbf{2})(5, -5)$,
    $(\mathbf{8}, \mathbf{1})(0, 0)$,
  \\
  &
  &
  &
  &
  &
  & $\{ (\bar{\mathbf{3}}, \mathbf{1})(2, 27) \}^{2}$,
  \\
  &
  &
  &
  &
  &
  & $(\mathbf{6}, \mathbf{1})(-68/11, 22)$,
  \\
  &
  &
  &
  &
  &
  & $(\bar{\mathbf{6}}, \mathbf{1})(68/11, -22)$
  \\ \hline
  %
  (9)
  & $\mathrm{Sp}(16)$
  & $\mathbf{544}$
  & $(\mathbf{1}, \mathbf{2})(3, -2)$,
  &
  & $(\mathbf{1}, \mathbf{1})(6, -4)$,
  & $\{ (\mathbf{1}, \mathbf{1})(0, 0) \}^{2}$,
  $(\mathbf{3}, \mathbf{1})(-8, 1)$,
  \\
  &
  &
  & $(\mathbf{1}, \mathbf{2})(-3, 2)$
  &
  & $(\mathbf{1}, \mathbf{2})(-3, 2)$,
  & $\{ (\mathbf{3}, \mathbf{1})(-2, -3) \}^{3}$,
    $\{ (\bar{\mathbf{3}}, \mathbf{1})(2, 3) \}^{2}$,
  \\
  &
  &
  &
  &
  & $(\bar{\mathbf{3}}, \mathbf{1})(-4, 7)$,
  & $(\bar{\mathbf{3}}, \mathbf{2})(5, 1)$,
    $(\mathbf{8}, \mathbf{1})(0, 0)$,
  \\
  &
  &
  &
  &
  & $(\bar{\mathbf{3}}, \mathbf{1})(2, 3)$,
  & $(\mathbf{6}, \mathbf{1})(2, 3)$,
    $(\bar{\mathbf{6}}, \mathbf{1})(-2, -3)$
  \\
  &
  &
  &
  &
  & $(\mathbf{3}, \mathbf{2})(1, -5)$
  &
  \\ \hline
  %
  (15a)
  & $\mathrm{SO}(13)$
  & $\mathbf{768}$
  & $(\mathbf{1}, \mathbf{2})(-3, 66)$,
  & $(\mathbf{3}, \mathbf{2})(1, 34)$,
  & $(\mathbf{1}, \mathbf{1})(-6, -36)$,
  & $(\mathbf{1}, \mathbf{2})(-9, 30)$,
    $(\mathbf{1}, \mathbf{2})(9, -30)$,
  \\
  &
  &
  & $(\mathbf{1}, \mathbf{2})(3, -66)$
  & $(\bar{\mathbf{3}}, \mathbf{1})(2, 100)$,
  & $(\mathbf{1}, \mathbf{2})(-9, 30)$,
  & $(\mathbf{3}, \mathbf{1})(4, -32)$,
    $(\bar{\mathbf{3}}, \mathbf{2})(-1, -34)$,
  \\
  &
  &
  &
  & $(\bar{\mathbf{3}}, \mathbf{1})(-4, 32)$,
  & $(\mathbf{1}, \mathbf{2})(9, -30)$,
  & $(\mathbf{3}, \mathbf{2})(-11, 38)$,
    $(\bar{\mathbf{3}}, \mathbf{2})(11, -38)$,
  \\
  &
  &
  &
  & $(\mathbf{1}, \mathbf{2})(-3, -102)$,
  & $(\mathbf{1}, \mathbf{2})(3, 102)$,
  & $(\mathbf{6}, \mathbf{1})(-4, 32)$,
    $(\bar{\mathbf{6}}, \mathbf{1})(4, -32)$,
  \\
  &
  &
  &
  &
    $(\mathbf{1}, \mathbf{1})(6, 36)$,
  & $(\bar{\mathbf{3}}, \mathbf{2})(-1, -34)$,
  & $(\mathbf{6}, \mathbf{2})(-1, -34)$,
    $(\bar{\mathbf{6}}, \mathbf{2})(1, 34)$,
  \\
  &
  &
  &
  &
    $(\mathbf{3}, \mathbf{3})(0, 8)$,
  & $(\mathbf{3}, \mathbf{1})(4, -32)$,
  & $(\mathbf{8}, \mathbf{1})(-4, 32)$,
    $(\mathbf{8}, \mathbf{1})(4, -32)$,
  \\
  &
  &
  &
  & $(\bar{\mathbf{3}}, \mathbf{3})(0, -8)$,
  & $(\mathbf{1}, \mathbf{1})(-6, -36)$
  & $(\mathbf{8}, \mathbf{2})(-1, -34)$,
    $(\mathbf{8}, \mathbf{2})(1, 34)$,
  \\
  &
  &
  &
  & $(\mathbf{1}, \mathbf{3})(-6, -36)$,
  &
  & $(\mathbf{3}, \mathbf{1})(4, -32)$,
    $(\bar{\mathbf{3}}, \mathbf{2})(-1, -34)$,
  \\
  &
  &
  &
  & $(\mathbf{1}, \mathbf{3})(6, 36)$,
  &
  & $(\mathbf{1}, \mathbf{1})(-6, -36)$,
    $(\mathbf{1}, \mathbf{2})(-9, 30)$,
  \\
  &
  &
  &
  & $(\mathbf{3}, \mathbf{1})(4, -32)$,
  &
  & $(\mathbf{1}, \mathbf{2})(9, -30)$,
    $(\bar{\mathbf{3}}, \mathbf{1})(2, 100)$,
  \\
  &
  &
  &
  & $(\bar{\mathbf{3}}, \mathbf{2})(-1, -34)$,
  &
  & $\{ (\mathbf{3}, \mathbf{2})(1, 34) \}^{5}$,
  \\
  &
  &
  &
  & $(\bar{\mathbf{3}}, \mathbf{2})(-7, 98)$
  &
  & $\{ (\bar{\mathbf{3}}, \mathbf{1})(-4, 32) \}^{2}$,
    $\{ (\mathbf{1}, \mathbf{1})(6, 36) \}^{2}$
  \\ \hline
  %
  (14)
  & $\mathrm{Sp}(14)$
  & $\mathbf{350}$
  & $(\mathbf{1}, \mathbf{2})(3, -2)$,
  &
  & $(\mathbf{1}, \mathbf{1})(6, -4)$,
  & $\{ (\mathbf{1}, \mathbf{1})(0, 0) \}^{2}$,
    $(\mathbf{3}, \mathbf{1})(-8, 1)$,
  \\
  &
  &
  & $(\mathbf{1}, \mathbf{2})(-3, 2)$
  &
  & $(\mathbf{1}, \mathbf{2})(-3, 2)$,
  & $\{ (\mathbf{3}, \mathbf{1})(-2, -3) \}^{3}$,
    $\{ (\bar{\mathbf{3}}, \mathbf{1})(2, 3) \}^{2}$,
  \\
  &
  &
  &
  &
  & $(\bar{\mathbf{3}}, \mathbf{1})(-4, 7)$,
  & $(\bar{\mathbf{3}}, \mathbf{2})(5, 1)$,
    $(\mathbf{8}, \mathbf{1})(0, 0)$,
  \\
  &
  &
  &
  &
  & $(\bar{\mathbf{3}}, \mathbf{1})(2, 3)$,
  & $(\mathbf{6}, \mathbf{1})(2, 3)$,
    $(\bar{\mathbf{6}}, \mathbf{1})(-2, -3)$
  \\
  &
  &
  &
  &
  & $(\mathbf{3}, \mathbf{2})(1, -5)$
  &
  \\ \hline
\end{tabular}
\end{center}
\end{table}


\vspace{2mm}  \noindent 
(a)
$S/R \, \textrm{(15a)} = \mathrm{G}_{2}/\mathrm{SU}(2) \times \mathrm{U}(1)$,
$G = \mathrm{Sp}(12)$, and $F = \mathbf{364}$.

We decompose $\mathrm{Sp}(12)$ as
\begin{eqnarray}
   \mathrm{Sp}(12)
 & \supset & \mathrm{Sp}(6) \times \mathrm{Sp}(6) \nonumber 
  \\
 & \supset & \mathrm{Sp}(6) \times \mathrm{Sp}(4) \times \mathrm{SU}(2)' \nonumber 
  \\
& \supset & \mathrm{SU}(3) \times \mathrm{Sp}(4) \times \mathrm{SU}(2)'  
            \times \mathrm{U}(1)_{\textrm{a}} \nonumber
  \\
& \supset & \mathrm{SU}(3) \times \mathrm{SU}(2) \times \mathrm{SU}(2) 
            \times \mathrm{SU}(2)' \times \mathrm{U}(1)_{\textrm{a}} \nonumber
  \\
& \supset & \mathrm{SU}(3) \times \mathrm{SU}(2) \times \mathrm{SU}(2)' 
            \times \mathrm{U}(1)_{\textrm{a}} \times \mathrm{U}(1)_{\textrm{b}}.
\end{eqnarray}
Accordingly the adjoint representation of $\mathrm{Sp}(12)$ is
decomposed as \cite{MckayPatera,Slansky:1981yr}
\begin{eqnarray}  
  \mathbf{78}
  &=&  (\mathbf{8}, \mathbf{1}, \mathbf{1})(0, 0)
  +  (\mathbf{1}, \mathbf{3}, \mathbf{1})(0, 0)
  + (\mathbf{1}, \mathbf{1}, \mathbf{3})(0, 0)
  + (\mathbf{1}, \mathbf{1}, \mathbf{1})(0, 0) \nonumber \\
& &+ (\mathbf{1}, \mathbf{1}, \mathbf{1})(0, 0) 
  + (\mathbf{6}, \mathbf{1}, \mathbf{1})(2, 0)
  + (\bar{\mathbf{6}}, \mathbf{1}, \mathbf{1})(-2, 0)
  + (\mathbf{3}, \mathbf{1}, \mathbf{2})(1, 0) \nonumber \\
 & &+ (\bar{\mathbf{3}}, \mathbf{1}, \mathbf{2})(-1, 0) 
  + (\mathbf{3}, \mathbf{2}, \mathbf{1})(1, 0) 
  + (\bar{\mathbf{3}}, \mathbf{2}, 1)(-1, 0)
  + (\mathbf{3}, \mathbf{1}, \mathbf{1})(1, 1) \nonumber \\
 & &+ (\bar{\mathbf{3}}, \mathbf{1}, \mathbf{1})(-1, -1)
  + (\mathbf{3}, \mathbf{1}, \mathbf{1})(1, -1)
  + (\bar{\mathbf{3}}, \mathbf{1}, \mathbf{1})(-1, 1) 
  + (\mathbf{1}, \mathbf{2}, \mathbf{1})(0, 1) \nonumber \\
 & &+ (\mathbf{1}, \mathbf{2}, \mathbf{1})(0, -1)
  + (\mathbf{1}, \mathbf{1}, \mathbf{2})(0, 1)
  + (\mathbf{1}, \mathbf{1}, \mathbf{2})(0, -1)
  + (\mathbf{1}, \mathbf{1}, \mathbf{1})(0, 2) \nonumber \\
 & &+ (\mathbf{1}, \mathbf{1}, \mathbf{1})(0, -2)
  + (\mathbf{1}, \mathbf{2}, \mathbf{2})(0, 0) \nonumber
  \\  \hspace{1.5em}
 & & (\mathrm{SU}(3), \mathrm{SU}(2), \mathrm{SU}(2)')
  (\mathrm{U}(1)_{\textrm{a}},\mathrm{U}(1)_{\textrm{b}}).
\end{eqnarray}
We take a linear combination of $\mathrm{U}(1)_{\textrm{a}}$ and
$\mathrm{U}(1)_{\textrm{b}}$, respecting the orthogonality of the two,
to obtain $\mathrm{U}(1)$ charges listed in
Table~\ref{so10_vector_and_spinor_with_u1u1}, at the row (15a) and the
columns ``Branch of $\mathbf{10}$'' and ``Branch of $\mathbf{16}$''.
We define
\begin{subequations}
\begin{align}
  Q_R & \equiv -x Q_{\textrm{a}} - y Q_{\textrm{b}}, \\
  Q_A & \equiv -2y Q_{\textrm{a}} + 3x Q_{\textrm{b}},
\end{align}
\end{subequations}
where $Q_{i}$s ($i \in \{ \textrm{a}, \, \textrm{b}, \, R, \, A \}$)
denote the charges of $\mathrm{U}(1)_{i}$.
Embedding $R$ in $\mathrm{SU}(2) \times \mathrm{U}(1)_{R}$, we obtain
the decomposition of the adjoint representation,
\begin{equation}
\begin{split}
  \mathbf{78}
  &
  = (\bar{\mathbf{8}}, \mathbf{1}, \mathbf{1})(0, 0)
  + (\bar{\mathbf{1}}, \mathbf{3}, \mathbf{1})(0, 0)
  + (\bar{\mathbf{1}}, \mathbf{1}, \mathbf{3})(0, 0)
  \\ & \quad
  + (\bar{\mathbf{1}}, \mathbf{1}, \mathbf{1})(0, 0)
  + (\bar{\mathbf{1}}, \mathbf{1}, \mathbf{1})(0, 0)
  \\ & \quad
  +(\bar{\mathbf{6}}, \mathbf{1}, \mathbf{1})(-2x, -4y)
  +(\bar{\mathbf{6}}, \mathbf{1}, \mathbf{1})(2x, 4y)
  \\ & \quad
  +(\bar{\mathbf{3}}, \mathbf{1}, \mathbf{2})(-x, -2y)
  +(\bar{\mathbf{3}}, \mathbf{1}, \mathbf{2})(x, 2y)
  \\ & \quad
  +(\bar{\mathbf{3}}, \mathbf{2}, \mathbf{1})(-x, -2y)
  +(\bar{\mathbf{3}}, \mathbf{2}, \mathbf{1})(x, 2y)
  \\ & \quad
  +(\bar{\mathbf{3}}, \mathbf{1}, \mathbf{1})(-x - y, -2y + 3x)
  \\ & \quad
  +(\bar{\mathbf{3}}, \mathbf{1}, \mathbf{1})(x + y, 2y - 3x)
  \\ & \quad
  +(\bar{\mathbf{3}}, \mathbf{1}, \mathbf{1})(-x + y, -2y - 3x)
  \\ & \quad
  +(\bar{\mathbf{3}}, \mathbf{1}, \mathbf{1})(x - y, 2y + 3x)
  \\ & \quad
  +(\bar{\mathbf{1}}, \mathbf{2}, \mathbf{1})(-y, 3x)
  +(\bar{\mathbf{1}}, \mathbf{2}, \mathbf{1})(y, -3x)
  \\ & \quad
  +(\bar{\mathbf{1}}, \mathbf{1}, \mathbf{2})(-y, 3x)
  +(\bar{\mathbf{1}}, \mathbf{1}, \mathbf{2})(y, -3x)
  \\ & \quad
  +(\bar{\mathbf{1}}, \mathbf{1}, \mathbf{1})(-2y, 6x)
  +(\bar{\mathbf{1}}, \mathbf{1}, \mathbf{1})(2y, 6x)
  \\ & \quad
  +(\bar{\mathbf{1}}, \mathbf{2}, \mathbf{2})(0, 0).
\end{split}
\label{dec78}
\end{equation}
We find that $y = \pm 2$ provides the SM Higgs doublet by comparing
the $\mathrm{U}(1)_R$ charges in the decomposition Eq.~(\ref{dec78})
with those in Table~\ref{so10_vector_and_spinor_with_u1u1}.
Further investigation shows that we can obtain the SM fermions as well
by taking $x = 1$ and $y = 2$.
The resulting field contents are summarized in
Table~\ref{table:SMGroup-decomp-RA}.
We can explicitly obtain appropriate $\mathrm{U}(1)_{Y}$ hypercharges
of the SM particles by taking another linear combination of
$\mathrm{U}(1)_{R}$ and $\mathrm{U}(1)_{A}$ as
\begin{subequations}
\begin{align}
  Q_Y & \equiv -\frac{6}{11} Q_R + \frac{7}{11} Q_A, \\
  Q_{\alpha}  & \equiv 19 Q_R + 2 Q_A,  
\end{align}
\end{subequations}
where $Q_{Y}$ and $Q_{\alpha}$ are the charges of $\mathrm{U}(1)_{Y}$
and $\mathrm{U}(1)_{\alpha}$, respectively.
We thereby obtain SM Higgs, SM fermions and other fermions listed as
in Table~\ref{table:SMGroup-decomp-Yalpha}.

\vspace{2mm} \noindent 
(b)
$S/R \, \textrm{(9)} = \mathrm{G}_{2} \times \mathrm{SU}(3) /
\mathrm{SU}(3) \times [\mathrm{SU}(2) \times \mathrm{U}(1)]$, $G =
\mathrm{Sp}(16)$, and $F = \mathbf{544}$.

We embed $R$ in subgroup $\mathrm{SU}(3)_{\textrm{b}} \times
\mathrm{SU}(2) \times \mathrm{U}(1)_R$ of $\mathrm{Sp}(16)$ according
to the decomposition
\begin{equation}
\begin{split}
  \mathrm{Sp}(16)
  & \supset
  \mathrm{Sp}(6)_{\textrm{a}}
  \times \mathrm{Sp}(6)_{\textrm{b}}
  \times \mathrm{Sp}(4)
  \\ 
  & \supset
  \mathrm{SU}(3)_{\textrm{a}}
  \times \mathrm{Sp}(6)_{\textrm{b}}
  \times \mathrm{Sp}(4)
  \times \mathrm{U}(1)_R
  \\ 
  & \supset
  \mathrm{SU}(3)_{\textrm{a}}
  \times \mathrm{SU}(3)_{\textrm{b}}
  \times \mathrm{Sp}(4)
  \\ & \quad
  \times \mathrm{U}(1)_R
  \times \mathrm{U}(1)_A 
  \\ 
  & \supset
  \mathrm{SU}(3)_{\textrm{a}}
  \times \mathrm{SU}(3)_{\textrm{b}}
  \times \mathrm{SU}(2)
  \times \mathrm{SU}(2)
  \\ & \quad
  \times \mathrm{U}(1)_R
  \times \mathrm{U}(1)_A. 
\end{split}
\end{equation}
The resulting field contents are summarized in
Table~\ref{table:SMGroup-decomp-RA}.
We explicitly obtain appropriate $\mathrm{U}(1)_Y$ hypercharges of the
SM particles by taking combination of $\mathrm{U(1)}_R$ and
$\mathrm{U(1)}_A$ as
\begin{subequations}
\begin{align}
  Q_Y & \equiv 3 Q_A - 2 Q_R, \\ 
  Q_{\alpha} & \equiv - 2 Q_A - 3 Q_R,
\end{align}
\end{subequations}
where $Q_{i}$s ($i \in \{ R, \, A, \, Y, \, \alpha \}$) denote the
charges of $\mathrm{U}(1)_{i}$.
We thereby obtain SM Higgs, SM fermions and other fermions listed in
Table~\ref{table:SMGroup-decomp-Yalpha}.

\vspace{2mm} \noindent  
(c)
$S/R \, \textrm{(15a)} = \mathrm{G}_{2} / \mathrm{SU}(2) \times \mathrm{U}(1)$,
$G = \mathrm{SO}(13)$, and $F = \mathbf{768}$ .

We decompose $\mathrm{SO}(13)$ as
\begin{eqnarray}
  \mathrm{SO}(13)
 & \supset &
  \mathrm{SU}(4)
  \times \mathrm{SO}(7) \nonumber 
  \\
  & \supset &
  \mathrm{SU}(4)
  \times \mathrm{SU}(2)''
  \times \mathrm{SU}(2)'
  \times \mathrm{SU}(2) \nonumber 
  \\
  & \supset &
  \mathrm{SU}(3)
  \times \mathrm{SU}(2)
  \times \mathrm{SU}(2)
  \times \mathrm{SU}(2)
  \times \mathrm{U}(1)_{\textrm{a}} \nonumber 
  \\
  & \supset &
  \mathrm{SU}(3)
  \times \mathrm{SU}(2)
  \times \mathrm{SU}(2)
  \times \mathrm{U}(1)_{\textrm{a}}
  \times \mathrm{U}(1)_{\textrm{b}},
\end{eqnarray}
where $\mathrm{SU}(2)'' \sim \mathrm{SO}(3)$ and $\mathrm{SU}(2)'
\times \mathrm{SU}(2) \sim \mathrm{SO}(4)$.
We obtain $\mathrm{U}(1)$ charges listed in
Table~\ref{so10_vector_and_spinor_with_u1u1} at the row of (15a) and
the column of ``Branch of $\mathbf{10}$'' and "Branch of
$\mathbf{16}$'' by taking a linear combination of
$\mathrm{U}(1)_{\textrm{a}}$ and $\mathrm{U}(1)_{\textrm{b}}$ as
\begin{align}
Q_R & \equiv \frac{3}{2} Q_{\textrm{b}} + \frac{1}{2} Q_{\textrm{a}} \\
Q_A & \equiv \frac{3}{2} Q_{\textrm{b}} -\frac{3}{2} Q_{\textrm{a}}, 
\end{align}
where $Q_i$ ($i \in \{ \textrm{a}, \, \textrm{b}, \, R, \, A \}$)
denote the charges of $\mathrm{U}(1)_{i}$.
Embedding $R$ in $\mathrm{SU(2)} \times \mathrm{U}(1)_{R}$, we obtain
the field contents summarized in Table~\ref{table:SMGroup-decomp-RA}.
We explicitly obtain appropriate $\mathrm{U}(1)_Y$ hypercharges of the
SM particles by taking another linear combination $\mathrm{U(1)}_R$
and $\mathrm{U(1)}_A$,
\begin{subequations}
\begin{align}
  Q_{Y} & \equiv -2 Q_{R} + Q_{A}, \\
  Q_{\alpha}  & \equiv 16 Q_{R} + 6 Q_{A},  
\end{align}
\end{subequations}
where $Q_{Y}$ and $Q_{\alpha}$ are the charges of $U(1)_{Y}$ and
$U(1)_{\alpha}$, respectively.
We thereby obtain SM Higgs, SM fermions and other fermions listed in
Table~\ref{table:SMGroup-decomp-Yalpha}.

\vspace{2mm} \noindent  
(d)
$S/R \, \textrm{(14)}=\mathrm{Sp}(6)/\mathrm{Sp}(4) \times \mathrm{U}(1)$,
$G = \mathrm{Sp}(14)$, and $F = \mathbf{350}$.

We decompose $\mathrm{Sp}(14)$ as
\begin{eqnarray}
    \mathrm{Sp}(14)
   & \supset &
    \mathrm{Sp}(10) \times \mathrm{Sp}(4) \nonumber
    \\
    & \supset &
    \mathrm{Sp}(6) \times \mathrm{Sp}(4)' \times \mathrm{Sp}(4) \nonumber 
    \\
    & \supset &
    \mathrm{SU}(3) \times \mathrm{Sp}(4)' \times \mathrm{Sp}(4)
    \times \mathrm{U}(1)_{\textrm{a}} \nonumber 
    \\
    & \supset &
    \mathrm{SU}(3) \times \mathrm{SU}(2) \times \mathrm{Sp}(4)
    \times \mathrm{U}(1)_{\textrm{a}} \times \mathrm{U}(1)_{\textrm{b}}.  
\end{eqnarray}
We obtain $\mathrm{U}(1)$ charges listed in
Table~\ref{so10_vector_and_spinor_with_u1u1} at the row of (14) and
the columns of ``Branch of $\mathbf{10}$'' and "Branch of
$\mathbf{16}$'' by taking a linear combinations of
$\mathrm{U}(1)_{\textrm{a}}$ and $\mathrm{U}(1)_{\textrm{b}}$ as
\begin{subequations}
\begin{align}
  Q_R & \equiv \frac{1}{2} (-9 Q_{\textrm{b}} + 2 Q_{\textrm{a}} ) \\
  Q_A & \equiv - Q_{\textrm{b}} -3 Q_{\textrm{a}},
\end{align}
\end{subequations}

where $Q_{i}$ ($i \in \{\textrm{a}, \, \textrm{b}, \, R, \, A \}$)
denote the charges of $\mathrm{U}(1)_i$.
Embedding $R$ in $\mathrm{Sp}(4) \times \mathrm{U}(1)_R$, we obtain
the resulting field contents summarized in
Table~\ref{table:SMGroup-decomp-RA}.
We explicitly obtain appropriate $\mathrm{U}(1)_Y$ hypercharges of the
SM particles by taking another linear combination of $\mathrm{U(1)}_R$
and $\mathrm{U(1)}_A$ as
\begin{subequations}
\begin{align}
  Q_Y & \equiv -\frac{2}{29} ( 5 Q_R + 21 Q_A ), \\
  Q_\alpha  & \equiv -\frac{2}{29} ( 14 Q_R - 5 Q_A ), 
\end{align}
\end{subequations}
where $Q_Y$ and $Q_{\alpha}$ are the charges of $\mathrm{U}(1)_Y$ and
$\mathrm{U}(1)_{\alpha}$.
We thereby obtain SM Higgs, SM fermions and other fermions listed in
Table~\ref{table:SMGroup-decomp-Yalpha}.

We find four candidates of $(S/R, G, F)$ which give the SM Higgs
doublet and at least one generation of the SM fermions in four
dimensions.
These models, however, generate numerous undesired fields that does
not appear in the particle spectrum of the SM as tabulated in
Table~\ref{table:SMGroup-decomp-Yalpha}.
These extra fields need to be eliminated to construct a realistic model
based on the candidates we found.

\section{Summary and discussions}
\label{sec:summary}

We analyzed gauge-Higgs unification models in a spacetime of the
dimensionality $D = 14$ under the scheme of the coset space
dimensional reduction and exhastively searched for the
phenomenologically acceptable models with the dimension of the fermion
representation less than 1024.
We first made a complete list of the fourteen-dimensional models by
determining the structure of the coset space $S/R$, the gauge group
$G$, and the representations $F$ of $G$ for fermions.
We obtained a full list of the possible cosets $S/R$ in
Table~\ref{table_10dim_coset_spaces} by requiring $\mathrm{dim} \, S/R
= 10$ and $\mathrm{rank} \, S = \mathrm{rank} \, R$.
The gauge groups $G$ are determined to have either complex or
pseudoreal representations (see Table~\ref{so10_vector_and_spinor_with_u1u1}), 
and to lead to one of the following two
types of gauge groups after the dimensional reduction to the
four-dimensional spacetime: the GUT-like gauge groups such as
$\mathrm{SO}(10) (\times \mathrm{U}(1))$ and $\mathrm{SU}(5) (\times
\mathrm{U}(1))$, or
the Standard-Model (SM)-like group which is
$\mathrm{SU}(3) \times \mathrm{SU}(2) \times \mathrm{U}(1) (\times
\mathrm{U}(1))$ (see Table~\ref{pairs}).
The representation $F$ of fermions are determined so that the matter
content of the SM emerges after the dimensional reduction.

We then analyzed the particle contents of the four-dimensional
theories that are induced from each of the sets $(S/R, G, F)$.
We found several interesting models in the GUT-like cases.

Among the interesting GUT-like models is the one with $H =
\mathrm{SO}(10) (\times \mathrm{U}(1))$, in which one or more
fermions of $\mathbf{16}$ representation, along with a number of
scalars of $\mathbf{10}$ representation, are derived in
four-dimensional theory.
A scalar of $\mathbf{10}$ can be interpreted as the electroweak
Higgs particle.
Two or more fermions of $\mathbf{16}$ in the models can account for
the generations of the fermions known in the particle spectra of the
SM.
The most interesting model in this point of view is the one for
\begin{math}
  S/R =
  \mathrm{SO}(7) \times \mathrm{SU}(3)
  / \mathrm{SO}(6) \times [\mathrm{SU}(2) \times \mathrm{U}(1)],
\end{math}
$G = \mathrm{SO}(20)$,
$F = \mathbf{512}$, and $H = \mathrm{SO}(10) \times \mathrm{U}(1)$.
Three fermions of $\mathbf{16}$ are obtained in this case,
suggesting the three generations of the fermions in the SM.
The $\mathrm{U}(1)$ charges associated to them imply a family
symmetry under this suggestion.

Similarly, a number of cases of $H = \mathrm{SU}(5) \times
\mathrm{U}(1)$ led to the models that induce fermions of
$\bar{\mathbf{5}}$ and $\mathbf{10}$ representations with a
scalar field of $\mathbf{5}$ representation.
Although the three sets of fermions are not obtained in these cases,
two of them are obtained for $G = \mathrm{Sp}(14)$,
\begin{math}
  S/R =
  \mathrm{Sp}(6) / \mathrm{Sp}(4) \times \mathrm{U}(1),
\end{math}
and $F = \mathbf{350}$, and can serve for the understanding of the
generations.

We also successfully constructed models for $H=\mathrm{SU}(3) \times
\mathrm{SU}(2) \times \mathrm{U}(1) \times \mathrm{U}(1)$, where Higgs
particle and a generation of the fermions are found. Many unwanted
fermions accompany them, however, and a mechanism to eliminate them is
necessary to build a realistic model.

In contrast, some of the GUT-like cases have only the desired
fermions.  It is worthwhile to analyze these models in further details.
An apparent challenge in the GUT-like cases, however, is the absence
of the Higgs particle which breaks the GUT gauge group down to the SM
gauge group.
We can employ the Hosotani mechanism, also known as the Wilson flux
breaking mechanism, to circumvent this difficulty.
More detailed analyses are necessary to examine if the models we found
interesting work in the phenomenological building of the models.

\section*{acknowledgments}
This work was supported in part by the Grant-in-Aid for the Ministry
of Education, Culture, Sports, Science, and Technology, Government of
Japan (No. 17740131, 18034001 and 19010485) and by MEC and FEDER (EC)
Grants No. FPA2005-01678 (T.~S.).

\end{document}